\documentstyle[12pt,epsf]{article}
\newcommand{\beq}{\begin{equation}}
\newcommand{\eeq}{\end{equation}}
\newcommand{\beqa}{\begin{eqnarray}}
\newcommand{\eeqa}{\end{eqnarray}}

\newcommand{\ve}{\varepsilon}

\begin{document}


\hfill KFA-IKP(Th)-1996-07

\hfill TK 96 24

\hfill hep-ph/9610275

\bigskip\bigskip

\begin{center}

{{\Large\bf Renormalization of the three flavor Lagrangian in \\[0.3cm]
    heavy baryon chiral perturbation theory}}\footnote{Work supported
    in part by Deutsche Forschungsgemeinschaft (grant ME 864-11/1)}

\end{center}

\vspace{.3in}

\begin{center}
{\large G. M\"uller$^\dagger$\footnote{email: mueller@pythia.itkp.uni-bonn.de},
 Ulf-G. Mei{\ss}ner$^\ddagger$\footnote{email: Ulf-G.Meissner@kfa-juelich.de}}

\bigskip

\bigskip

$^\dagger${\it Universit\"at Bonn, Institut f{\"u}r Theoretische Kernphysik\\
Nussallee 14-16, D-53115 Bonn, Germany}\\

\bigskip

$^\ddagger${\it Forschungszentrum J\"ulich, Institut f\"ur Kernphysik 
(Theorie)\\ D-52425 J\"ulich, Germany}

\end{center}

\vspace{.7in}

\thispagestyle{empty} 

\begin{abstract}
The complete renormalization of the generating functional for Green
functions of  quark currents between one--baryon  states
in three flavor heavy baryon chiral perturbation theory is performed 
to order $q^3$. As an example, we study the kaon loop induced
divergences in neutral pion photoproduction off protons.
\end{abstract}

\vspace{1in}

\centerline{ACCEPTED FOR PUBLICATION IN {\it NUCL. PHYS.} {\bf B}}

\vfill


\section{Introduction}

Chiral perturbation theory (CHPT) is the effective field theory of the
standard model at energies below the scale of spontaneous chiral
symmetry breaking QCD is believed to undergo. The basic degrees of
freedom are the eight (almost) massless pseudoscalar Goldstone bosons.
In the framework of the non--linear realization of the chiral
symmetry, it is also straightforward to implement matter fields (like
e.g. baryons) in the pertinent effective field theory \cite{wein68} 
\cite{ccwz}. The first systematic discussion for the two flavor
sector, i.e. the pion--nucleon system, implementing the ideas of a
chiral power counting \cite{wein79}, was performed in ref.\cite{gss}. However,
treating the nucleons as relativistic spin--1/2 (Dirac) fields does
not allow for a one--to--one correspondence between the expansion in
small momenta and quark masses on one  and the pion loop expansion
on the other side. As pointed out in ref.\cite{jm}, this short--coming
can be overcome if one makes use of methods borrowed from HQEFT,
namely to consider the baryons as extremely heavy, static
sources. This is called heavy baryon CHPT (HBCHPT).
 One considers a particular frame in which any baryon field
can be characterized by a four--velocity $v_\mu$.
 This allows to transform the troublesome baryon mass term in
the propagator into a string of $1/m$ suppressed meson--baryon
interaction vertices. A more formal approach based on path integral
methods was proposed in ref.\cite{bkkm}. In that scheme, it is
particularly simple to systematically construct all $1/m$ suppressed
vertices with fixed and with free coupling constants. Stated
differently, Lorentz invariance is automatically ensured since one
starts from the fully relativistic pion--nucleon Lagrangian  to
perform the frame--dependent decomposition of HBCHPT. To one loop
order, divergences appear. Some of these were treated e.g. in
\cite{bkkm}. A systematic treatment of the leading divergences of the
generating functional for Green functions of quark currents between
one--nucleon states was given in ref.\cite{ecker}. This allows for a
chiral invariant renormalization of {\it all} two--nucleon Green
functions of the pion--nucleon system to order $q^3$ in the
low--energy expansion, were $q$ denotes a small momentum or Goldstone
boson mass. This complete divergence structure listed in \cite{ecker}
was e.g. heavily used as a check in the calculation of the reaction
$\pi N \to \pi \pi N$ to ${\cal O}(q^3)$ \cite{bkmpipin}. Furthermore, in
ref.\cite{eckmoj} nucleon field transformations were used to bring the
renormalized pion--nucleon Lagrangian in standard form, i.e. all
finite and infinite one--particle--irreducible vertices were obtained.
In that paper, it was also stressed that one has to consistently construct
the meson and the meson--baryon Lagrangians since otherwise one is left
with unwanted divergences in the reducible functional diagrams at 
order $q^3$ proportional to the equations of motion for the meson fields.
Further applications to the 
pion--nucleon system to one loop accuracy are summarized in the 
reviews~\cite{ulfrev}~\cite{bkmrev}~\cite{eckerrev}.

The situation is very different in SU(3). Although the original heavy
fermion approach was formulated for three flavors \cite{jm}, most of
the corresponding calculations of baryon masses, magnetic moments,
hyperon polarizabilities and decays and so on are either finite to order $q^3$
or only the leading non--analytic pieces (at a scale of 1~GeV) were
accounted for. The only works in which renormalization at order $q^3$
has been performed are concerned with the kaon--nucleon interaction
\cite{rho}, kaon photo-- and electroproduction \cite{sven} and strong
and electromagnetic decays of the decuplet (where the EFT is extended
to include the spin--3/2 fields) \cite{bss}. It is our aim to work out
{\it all} leading  divergences in the generating functional of  the three
flavor meson--baryon interaction, thus extending the work of Ecker
\cite{ecker} making use of the same methods. The main difference to
the two flavor case is the richer structure of the corresponding Lie
algebra, thus leading to much more allowed terms at a given order in
the chiral Lagrangian.  Stated differently, the main difference
between the two calculations lies in the fact that the nucleons are 
in the fundamental representation of SU(2), while the baryons are in 
the adjoint representation of SU(3).This leads to some algebraic
consequences for the construction of the one-loop generating
functional to be discussed below. We remark that all terms to order $q^3$ in
relativistic SU(3) baryon CHPT were enumerated in \cite{krause}. In
that paper, mass and wavefunction renormalization was
performed. Obviously, only a subset of the operators listed in
\cite{krause} contains  divergences. 

The manuscript is organized as follows. In section~\ref{sec:HBCHPT}
we review the path integral formalism of heavy baryon CHPT extended to
the three flavor case. In section~\ref{sec:genone}, we work out the
generating functional to one loop, i.e. to order $q^3$. Here and in
what follows, our work closely parallels the one of ref.\cite{ecker}.
However, we give a more detailed exposition of the method.  Some
formalism related to the heat kernel technique is spelled out in 
section~\ref{sec:heat}. Then, in section~\ref{sec:tadpole} we work out
the renormalization of the irreducible tapdole graph. The much more
involved renormalization of the irreducible self--energy graph is
spelled out in section~6. In section~7, we write down the full
counterterm Lagrangian at order $q^3$ and tabulate the pertinent
operators and their $\beta$--functions. This table constitutes the
main result of this investigation. Section~8 contains a sample
calculation for $\pi^0$ photoproduction off protons. Here, kaon loops
lead to a $q^3$ divergence. We give straightforward Feynman diagram
evaluation of this divergence and show how to use table~1. A summary
and a discussion of the various checks on our calculation is given in
section~9. The appendices contain sufficiently detailed technicalities
to check the calculation at various intermediate steps.


\section{Heavy baryon formalism: path integral approach}
\label{sec:HBCHPT}

The interactions of the Goldstone bosons with the ground state baryon
octet states are severely constrained by chiral symmetry. The
generating functional for Green functions of quark currents 
between single baryon states, $Z[j,\eta, \bar \eta$], is defined via
\beq
\exp \, \bigl\{ i \, Z[j,\eta, \bar \eta] \bigr\} 
= {\cal N} \int [du] [dB] [d{\bar B}] \, \exp 
\, i \biggl[ S_M + S_{MB} + \int d^4x <\bar \eta B> +<\bar B \eta>
\biggr] \, \, \, ,
\label{defgenfun}
\eeq
with $S_M$ and $S_{MB}$ denoting the mesonic and the meson--baryon
effective action, respectively, to be discussed below. $\eta$ and
$\bar \eta$ are fermionic sources coupled to the baryons and $j$
collectively denotes the external fields of vector ($v_\mu$), 
axial--vector ($a_\mu$), scalar ($s$) and pseudoscalar ($p$) type. 
These are coupled in the standard chiral
invariant manner. In particular, the scalar source contains the quark
mass matrix ${\cal M}$, $s(x) = {\cal M} + \ldots$. Traces in flavor
space are denoted by $<...>$. The underlying effective Lagrangian
can be decomposed into a purely mesonic ($M$) and a meson--baryon ($MB$)
part as follows (we only consider processes with exactly one baryon in
the initial and one in the final state)
\beq
{\cal L}_{\rm eff} = {\cal L}_M + {\cal L}_{MB}
\eeq
subject to the following low--energy expansions
\beq
{\cal L}_M =  {\cal L}_M^{(2)} + {\cal L}_M^{(4)} + \ldots  \, \, ,
\quad {\cal L}_{MB} =  {\cal L}_{MB}^{(1)} + {\cal L}_{MB}^{(2)}
+ {\cal L}_{MB}^{(3)} + \ldots \,
\eeq
where the superscript denotes the chiral dimension. The lowest order
meson Lagrangian takes the form \cite{gl85}
\beqa
 {\cal L}_M^{(2)} &=& \frac{F_0^2}{4} \, < u_\mu u^\mu + \chi_+> \,\, ,
\nonumber \\
u_\mu   &=& i [ u^\dagger (\partial_\mu-ir_\mu) u -
                u (\partial_\mu-il_\mu) u^\dagger ] \nonumber \\
        &=& i [ \xi^\dagger_R(\partial_\mu-ir_\mu)\xi_R -
              \xi^\dagger_L(\partial_\mu-il_\mu)\xi_L ]  \,\, ,
              \nonumber \\
\chi_{\pm} &=& u^\dagger \chi u^\dagger \pm u \chi^\dagger u \, \, .
\label{Lmeson}
\eeqa
The pseudoscalar Goldstone fields ($\phi = \pi, K, \eta$) are collected in
the  $3 \times 3$ unimodular, unitary matrix $U(x)$, 
\begin{equation}
 U(\phi) = u^2 (\phi) = \exp \lbrace i \phi / F_0 \rbrace
\label{U}
\end{equation}
with $F_0$ the pseudoscalar decay constant (in the chiral limit), and
\begin{eqnarray}
 \phi =  \sqrt{2}  \left(
\matrix { {1\over \sqrt 2} \pi^0 + {1 \over \sqrt 6} \eta
&\pi^+ &K^+ \nonumber \\
\pi^-
        & -{1\over \sqrt 2} \pi^0 + {1 \over \sqrt 6} \eta & K^0
        \nonumber \\
K^-
        &  \bar{K^0}&- {2 \over \sqrt 6} \eta  \nonumber \\} 
\!\!\!\!\!\!\!\!\!\!\!\!\!\!\! \right) \, \, \, \, \, . 
\end{eqnarray}
Under SU(3)$_L \times$SU(3)$_R$, $U(x)$ transforms as $U \to U' =
LUR^\dagger$, with $L,R \in$ SU(3)$_{L,R}$. Furthermore, $\xi_L$,
$\xi_R$ are elements of the chiral coset space 
SU(3)$_L \times$SU(3)$_R$/SU(3)$_V$ with $U(\phi) = \xi_R
(\phi)\xi_L^\dagger (\phi) 
$. The more familiar choice for the $\xi_{L,R}$ is $u
(\phi) =\xi_R = \xi^\dagger_L$, compare Eq.(\ref{U}). The external fields
appear in Eq.(\ref{Lmeson}) in the following chiral invariant
combinations,
\beq
r_\mu = v_\mu +a_\mu \, \, , \quad
l_\mu = v_\mu -a_\mu \, \, , \quad
\chi = 2 B_0 \,(s+ip) \, \, ,
\eeq
and $B_0$ is related to the quark condensate in the chiral limit,
$B_0 = |<0|\bar q q|0>|/F_0^2$. We adhere to the standard chiral
counting,
\beqa
{\cal O}(1) &:& U \, , \, u \, , \, \xi_L \, , \, \xi_R \, \, ,
\nonumber\\
{\cal O}(q) &:& \partial_\mu \, , \, l_\mu \, , \, r_\mu \, , \, u_\mu \, \, ,
\nonumber\\
{\cal O}(q^2) &:& s \, , \, p \, , \, F^{\mu \nu}  \, ,
\eeqa
with $q$ denoting a small momentum or meson mass.

The effective meson--baryon Lagrangian starts with terms of dimension one,
\beqa
{\cal L}_{MB}^{(1)} & = & < \bar B \, [ \, i \nabla\!\!\!\!/ \, , \, B] \, > \,
- \, m < \bar B  \, B> \nonumber  \\
&&
+ {D \over 2} \, < \bar B \, \{u\!\!\!/ \gamma_5 , B \}\,>
+ {F \over 2} \, < \bar B \, [ u\!\!\!/ \gamma_5 , B ] \,> \,\,\, ,
\label{LMB1}
\eeqa
with $m$ the average octet mass in the chiral limit.
The $3\times 3$ matrix $B$ collects the baryon octet, 
\begin{eqnarray}
B  =  \left(
\matrix  { {1\over \sqrt 2} \Sigma^0 + {1 \over \sqrt 6} \Lambda
&\Sigma^+ &  p \nonumber \\
\Sigma^-
    & -{1\over \sqrt 2} \Sigma^0 + {1 \over \sqrt 6} \Lambda & n
    \nonumber \\
\Xi^-
        &       \Xi^0 &- {2 \over \sqrt 6} \Lambda \nonumber \\} 
\!\!\!\!\!\!\!\!\!\!\!\!\!\!\!\!\! \right)  \, \, \, .
\end{eqnarray}
Under $SU(3)_L \times SU(3)_R$, $B$  transforms as any matter field,
\begin{equation} 
B \to B' = K \, B \,  K^\dagger
 \, \, \, ,
\end{equation}
with $K(U,L,R)$ the compensator field representing an element of the
conserved subgroup SU(3)$_V$. $\nabla_\mu$ denotes the covariant
derivative,
\beq
[\nabla_\mu , B] = \partial_\mu \, B + [ \, \Gamma_\mu , B \,]
\label{nablaB}
\eeq
and $\Gamma_\mu$ is the chiral connection,
\beqa
\Gamma_\mu & = & \frac{1}{2}\, [ u^\dagger (\partial_\mu-ir_\mu) u +
u (\partial_\mu-il_\mu) u^\dagger ] \nonumber \\
 & = & \frac{1}{2}\,
 [ \xi^\dagger_R(\partial_\mu-ir_\mu)\xi_R +
 \xi^\dagger_L(\partial_\mu -il_\mu)\xi_L ]  \,\,\, .
\eeqa
Note that the first term in
Eq.(\ref{LMB1}) is of dimension one since $[ i \nabla \!\!\!\!/ \, , \,
B ] - m \, 
B = {\cal O}(q)$ \cite{gss}. The lowest order
meson--baryon Lagrangian contains two axial--vector coupling
constants, denoted by $D$ and $F$. The dimension two and three terms
have been enumerated by Krause \cite{krause}. Treating the baryons as
relativistic spin--1/2 fields, the chiral power counting is no more
systematic due to the large mass scale $m$, $\partial_0 \, B 
\sim m \, B \sim \Lambda_\chi \, B$. This problem can be overcome in
the heavy mass formalism proposed in \cite{jm}. 
We follow here the path integral approach  developed in \cite{bkkm}.
Defining velocity--dependent spin--1/2 fields by a particular choice 
of Lorentz frame and decomposing the fields into their velocity
eigenstates (sometimes called 'light' and 'heavy' components),
\beqa
H_v (x) &=& \exp \{ i m v \cdot x \} \, P_v^+ \, B(x) \,\, , \nonumber \\
h_v (x) &=& \exp \{ i m v \cdot x \} \, P_v^- \, B(x) \,\, , \nonumber \\
P_v^\pm &=& \frac{1}{2} (1 \pm v \!\!\!/) \,\, , \,\,\, v^2 = 1 \,\,\, ,
\label{Bheavy}
\eeqa
the mass dependence is shuffled from the fermion propagator into a
string of $1/m$ suppressed interaction vertices. In this basis, the 
three flavor meson--baryon action takes the form
\beq
S_{MB} = \int d^4x \, \biggl\{ \bar{H}_v^a \, A^{ab} H^b_v 
- \bar{h}_v^a \, C^{ab} h^b_v + \bar{h}_v^a \, B^{ab} H^b_v
+ \bar{H}_v^a \, \gamma_0 \,  {B^{ab}}^\dagger \, \gamma_0 \, h^b_v 
\biggr\}\,\,\, ,
\eeq
with $a,b= 1, \ldots ,8$ flavor indices. The 8$\times$8 matrices $A$,
$B$ and $C$ admit low energy expansions, in particular
\beqa
A^{ab}       &=& A^{ab}_{(1)}+ A^{ab}_{(2)} + A^{ab}_{(3)} + \ldots
\nonumber \\
A^{ab}_{(1)} &=& < \, {\lambda^a}^\dagger \, [i v \cdot \nabla \, , \lambda^b
] > + D \, < \, {\lambda^a}^\dagger \, \{S \cdot u, \lambda^b\} \, >
     + F \, < \, {\lambda^a}^\dagger \, [ S \cdot u, \lambda^b ] \, >
\, \, , \nonumber \\
A^{ab}_{(2)} &=&  b_{D/F} \, <{\lambda^a}^\dagger \, (\chi_+,
\lambda^b)_\pm> + b_0 \,  <{\lambda^a}^\dagger \, \lambda^b><\chi_+>
\nonumber \\
&+& b_{1,2,3} \, <{\lambda^a}^\dagger \, ( u_\mu , (u^\mu ,
\lambda^b )_\pm )_\pm>
+ \,b_{4,5,6} \, <{\lambda^a}^\dagger \, ( v \cdot u , (v \cdot u ,
\lambda^b )_\pm )_\pm> \nonumber \\
&+& b_7 \,  <{\lambda^a}^\dagger \, \lambda^b><u^2> + \, b_8 \,
 <{\lambda^a}^\dagger \, \lambda^b><(v \cdot u)^2 > \nonumber \\
&+& 2i v_\rho \epsilon^{\rho \mu \nu \sigma} S_\sigma \, \biggl\{
 b_{9} \, <{\lambda^a}^\dagger \, u_\mu><u_\nu \, \lambda^b> 
+  b_{10,11} \, <{\lambda^a}^\dagger \, ([u_\mu , u_\nu ],
\lambda^b)_\pm>  \nonumber \\
&&\,\,\,\,\,\,\,\,\,\,\,\,\,\,\,\,\,\,\,\,\,\,\,\,\,\,\,\,\,\,\,\,\,\,
 \, -i \, b_{12,13} \, <{\lambda^a}^\dagger \, (F_{\mu\nu}^+ ,
\lambda^b)_\pm> \biggr\} \nonumber \\
B^{ab}_{(1)} &=&  < \, {\lambda^a}^\dagger \,[ i 
{\nabla}^{\!\perp}\!\!\!\!\!\!\!/ \,\, ,  \lambda^b ] \, > - \frac{D}{2} \, 
< \, {\lambda^a}^\dagger \, \{v \cdot u \gamma_5, \lambda^b\} \, >
- \frac{F}{2} \, < \, {\lambda^a}^\dagger \, [ v \cdot u \gamma_5, 
\lambda^b ] \, > \,\, , \nonumber \\
C^{ab}_{(1)} &=& A^{ab}_{(1)} + 2 m \, \delta^{ab}
\eeqa
where $\lambda^a$ denotes the  SU(3) matrices in the {\it physical}
basis,
\beq
O(x) = O^a(x) \, \lambda^a \,\, , \quad O=\{B,H_v,h_v\} \,\, , \quad
<{\lambda^a}^\dagger \lambda^b> = \delta^{ab} \,\,\, ,
\eeq
and $S^\mu$ is the covariant spin--operator {\`a} la Pauli--Lubanski,
\beq
S^\mu = \frac{i}{2} \, \gamma_5 \, \sigma^{\mu \nu} \, v_\nu \,\, \, ,
\eeq
subject to the constraint $S \cdot v = 0$, with
\beq
\label{spinrel}
[S_\mu,S_\nu] = i \, \epsilon_{\mu\nu\sigma\rho} \, v^\sigma \, S^\rho
\,\, , \,\, \{S_\mu,S_\nu\} = \frac{1}{2} ( v_\mu v_\nu - g_{\mu\nu} )
\,\, , \, \, S^2 =- \frac{d-1}{4} \,\, .
\eeq
Furthermore,
${\nabla}^\perp\!\!\!\!\!\!\!\!/ 
\,\,\,\,\, = \gamma^\mu (v_\mu v_\nu - g_{\mu\nu})  \nabla^\nu $.
We have introduced the compact notation
\beq
D/F< \ldots(Q,O)_\pm \ldots > \equiv D < \ldots \{Q,O\} \ldots > +
F < \ldots [Q,O] \ldots > 
\eeq
for any 3$\times$3 matrices $Q$ and $O$. Notice that the
(anti)commutators $(..)_\pm$ only act in flavor space and that spin--matrices
appearing in the operators have all to be taken to the left in the
appropriate order. To reduce the number of terms in $A^{ab}_{(2)}$, we
have made use of the Cayley-Hamilton relation for traceless matrices
$X = \{ u_\mu , v \cdot u \} $ 
\beq
\label{CH}
<{\lambda^a}^\dagger \, X >< X \, \lambda^b> =
<{\lambda^a}^\dagger \, \{ X^2 , \lambda^b \}> +
<{\lambda^a}^\dagger \, X \, \lambda^b \, X > -
\frac{1}{2}<{\lambda^a}^\dagger \, \lambda^b> <X^2> \,\, .
\eeq
We also have $F_{\mu\nu}^+ = u F_{\mu\nu}^L u^\dagger + u^\dagger
F_{\mu\nu}^R u$, with $ F_{\mu\nu}^{L,R}$ the field strength tensors
related to $l_\mu$ and $r_\mu$, respectively.
Similarly, we split the
baryon source fields $\eta (x)$ into velocity eigenstates,
\beqa
R_v (x) &=& \exp \{ i m v \cdot x \} \, P_v^+ \, \eta(x) \,\, , \nonumber \\
\rho_v (x) &=& \exp \{ i m v \cdot x \} \, P_v^- \, \eta(x) \,\, , 
\label{sourceheavy}
\eeqa
and shift variables
\beq
h_v^{a'} = h_v^a - (C^{ac})^{-1} \, ( B^{cd} \, H_v^d + \rho_v^c \, )
\,\, \, ,
\eeq
so that the generating functional takes the form
\beq
\exp[iZ] = {\cal N} \, \Delta_h \, \int [dU][dH_v][d\bar{H}_v] \, \exp \bigl\{
iS_M + i S_{MB}' \, \bigr\}
\label{Zinter}
\eeq
in terms of the meson--baryon action $S_{MB}'$,
\beqa
S_{MB}' &=& \int d^4x \, \bar{H}_v^a \bigl( A^{ab} + \gamma_0
[B^{ac}]^\dagger \gamma_0 [C^{cd}]^{-1} B^{db} \, \bigr) H_v^b
\nonumber \\
&+& \bar{H}_v^a \bigl( R_v^a + \gamma_0 [B^{ac}]^\dagger \gamma_0 
[C^{cd}]^{-1} \rho_v^d \bigr) + \bigr( \bar{R}_v^a + \bar{\rho}_v^c
[C^{cb}]^{-1} B^{ba} \bigr) H_v^a \, \, .
\eeqa        
The determinant $\Delta_h$ related to the 'heavy' components is
identical to one as first noted in \cite{mannel}. This can be
understood from the observation that
this determinant is related to anti--particle propagation which
decouples completely to the order we are working. The generating
functional is thus entirely expressed in terms of the Goldstone bosons
and the 'light' components of the spin--1/2 fields. The action is,
however, highly non--local due to the appearance of the inverse of 
the matrix $C$. To render it local, one now expands $C^{-1}$ in powers
of $1/m$, i.e. in terms of increasing chiral dimension,
\beqa
[C^{ab}]^{-1} &=& \frac{1}{2m} \, \sum_{n=0}^\infty \, (-1)^n \, \biggl(
\biggl[ \, \frac{C-2m}{2m} \biggr]^{ab} \, \biggr)^n \,\, , \nonumber\\
 &=& \frac{\delta^{ab}}{2m} - \frac{A^{ab}_{(1)}}{(2m)^2} + {\cal
   O}(q^2) \,\,\, .
\label{Cinvexp}
\eeqa
For calculating the higher order corrections to $C^{-1}$ not exhibited
in Eq.(\ref{Cinvexp}), it is advantageous to make use of the completeness
relation
\beq
<{\lambda^a}^\dagger \, O \, \lambda^b> <{\lambda^b}^\dagger \, Q \, 
\lambda^c> = <{\lambda^a}^\dagger \, O Q \, \lambda^c> - \frac{1}{3}
<{\lambda^a}^\dagger \, O> < Q \, \lambda^c> \,\, , \quad O,Q \in {\rm
  SU(3)}
\label{comprel}
\eeq
to bring the pertinent expressions in a more compact form.
To any finite power in $1/m$, one can now perform the integration of
the 'light' baryon field components $N_v$ by again completing the
square,
\beqa
H_v^{a'} &=& [T^{ac}]^{-1} \, \bigl( R_v^c + \gamma_0 \, [B^{cd}]^\dagger
\, \gamma_0 \, [C^{db}]^{-1} \, \rho_v^b \, \bigr) \nonumber \\
T^{ab} &=& A^{ab} + \gamma_0 \, [B^{ac}]^\dagger \, \gamma_0 \,
[C^{cd}]^{-1} \, B^{db} \,\,\, .
\eeqa
Notice that the second term in the expression for $T^{ab}$ only starts
to contribute at chiral dimension two (and higher). To be more
precise, we give the chiral expansion of $T^{ab}$ up to and including
all terms of order $q^3$,
\beqa
T^{ab} &=& A^{ab}_{(1)} + A^{ab}_{(2)} + A^{ab}_{(3)} + \frac{1}{2m}
\, \gamma_0 \, [B^{ac}]^\dagger_{(1)} \, \gamma_0 \, B^{cb}_{(1)}
\nonumber \\
&+& \frac{1}{2m} \biggl( \, \gamma_0 \, [B^{ac}]^\dagger_{(1)} \, 
\gamma_0 \, B^{cb}_{(2)} +  \gamma_0 \, [B^{ac}]^\dagger_{(2)} \, 
\gamma_0 \, B^{cb}_{(1)} \biggr) \nonumber \\
&-& \frac{1}{(2m)^2} 
\, \gamma_0 \, [B^{ac}]^\dagger_{(1)} \, \gamma_0 \, C_{(1)}^{cd} \,
B^{db}_{(1)} + {\cal O}(q^4) \,\,\, .
\eeqa
We thus arrive at
\beq
\exp[iZ] = {\cal N}' \, \int [dU] \, \exp \bigl\{ iS_M + i Z_{MB} \, \bigr\}
\,\,\, ,
\label{Zfinal}
\eeq
with ${\cal N}'$ an irrelevant normalization constant. The generating 
functional has thus been reduced to the purely mesonic
functional. $Z_{MB}$ is given by
\beqa
Z_{MB} = - \int d^4x &\biggl\{& 
 \bar{\rho}^a_v \, \bigl( [C^{ac}]^{-1} \, B^{cd}
\, [T^{de}]^{-1} \, \gamma_0 \, [B^{ef}]^\dagger \, \gamma_0 [C^{fb}]^{-1} -
[C^{ab}]^{-1} \, \bigr) \, \rho_v^b
\nonumber \\
&+& 
 \bar{\rho}^a_v \, \bigl( \, [C^{ac}]^{-1} \, B^{cd} \, [T^{db}]^{-1} 
\, \bigr) \, R_v^b + \bar{R}_v^a \, \bigl( [T^{ac}]^{-1} \, \gamma_0 \,
[B^{cd}]^\dagger \, \gamma_0 \, [C^{db}]^{-1} \, \bigr) \, \rho_v^b
\nonumber \\
&+&  \quad \bar{R}_v^a \, [T^{ab}]^{-1} \, R_v^b 
\quad \biggr\} \,\,\,\,\, . 
\label{ZMB}
\eeqa
At this point, it is important to determine the chiral dimension with which
the various terms in Eq.(\ref{ZMB}) start to contribute. The first
term in the first line obviously only starts at order $q^3$ since both
$B^{ab}$ and $T^{ab}$ start out at ${\cal O}(q)$. The second term in
the first line is more tricky. As shown in Eq.(\ref{Cinvexp}), $C^{-1}$
starts with terms of order one. However, these are exactly the
contributions from anti--particle propagation, i.e. $C^{-1}$ is
related to the anti--baryon propagator as shown in
Eq.(\ref{Antilowest}). While the baryon propagator 
does not contain the mass any more, the anti--baryon propagator picks
up exactly the factor $2m$ which is nothing but the gap between the
particle and the anti--particle sectors. A more formal argument
is given below. The terms in the second line of Eq.(\ref{ZMB})
are of order $q^2$ (and higher) and the term in the third line starts
out at ${\cal O}(q)$. Consequently, to order $q^3$, only this last 
term in Eq.(\ref{ZMB}) generates the Green functions related to the
'light' fields.  This also means that extending the calculation one
order further, i.e. to ${\cal O}(q^4)$, will lead to a much more
complicated structure than the one discussed here. We hope to come
back to this point at a later stage. Let us return to the question of
anti--particle propagation. We conjecture that the propagator
$C^{-1}(x)$ describes backward propagation along the time--like vector
$v$. This implies that fermion loops in the small momentum expansion
can never be closed in coordinate space. Consider first the lowest
order. With $C^{(1)}=(i v \cdot \partial + 2m)$ one gets the
free propagator 
\beqa
S^{(1)}(x) &=& \int \frac{d^4 k}{(2\pi)^4} \, \frac{1}{v \cdot k + 2m - i
  \epsilon} \, e^{-ik x} \nonumber \\
&=& i \, e^{2mi \, v \cdot x} \, \Theta (-v \cdot x) \, \int \, 
\frac{d^4 k}{(2\pi)^3} \, \delta(k \cdot v) \,e^{-ik x}\,\, ,
\label{Antilowest}
\eeqa
or in the rest--frame of the antiparticle,
\beq
S^{(1)}(x) = i \, e^{ 2 i m \, t} \, \Theta(-t) \, \delta( \vec r \,)
\eeq
which shows that the anti--baryon is sharply located in space. 
At next order,
\beq
C^{(2)} = - A_{(2)} \,\, ,
\eeq
and the free propagator is given by Eq.(\ref{Antilowest}) which shows
that $C_{(1)} + C_{(2)}$ corresponds to an antiparticle. On the other hand,
for the lowest two orders  of the 'light' fields we find 
\beqa
S^{(1)}(x) &=& \int \frac{d^4 k}{(2\pi)^4} \, \frac{1}{v \cdot k  + i
  \epsilon} \, e^{-ik x} \nonumber \\
&=& -i \,  \, \Theta (v \cdot x) \, \int \, 
\frac{d^4 k}{(2\pi)^3} \, \delta(k \cdot v) \,e^{-ik x}\,\, ,
\label{Particlelowest}
\eeqa
and with the contribution to the next order  
\beq
A^{(2)} = \frac{1}{2m} (v \cdot \partial)^2
-\frac{1}{2m} \partial^2 \,\, ,
\eeq
the corresponding free propagator is given by
\beqa
S^{(2)}(x) &=& \int \frac{d^4 k}{(2\pi)^4} \, \frac{1}{v \cdot k
-(v \cdot k)^2/2m + k^2/2m + i  \epsilon} \, e^{-ik x} \nonumber \\
&=& -i \,  \, \Theta (v \cdot x) \, \int \,
\frac{d^4 k}{(2\pi)^3} \, \delta(k \cdot v) \,e^{-ik x}
e^{ i  k^2(v \cdot x)/2m}\,\, ,
\eeqa
which again shows that the particle propagates forward in times. It
is, however, no longer sharply located at the point $\vec r\,$ in
coordinate space but rather shows some spread. Higher orders can be
treated along the same lines, i.e. to any finite order in $1/m$ the
particle and anti--particle sectors do not interact (see also Ecker's
proof that there are no closed baryon loops \cite{eckerc}).

To end this section, we give the chiral dimension $D$ for processes
with exactly one baryon line running through the pertinent Feynman
diagrams,
\beq
D = 2L +1 + \sum_{d=4,6,\ldots} (d-2) \, N_d^M + \sum_{d=2,3,\ldots} 
(d-1) \, N_d^{MB} \ge 2L+1
\label{chdim}
\eeq
with $L$ denoting the number of (meson) loops, and $N_d^M$ ($N_d^{MB}$) 
counts the number of mesonic (meson--baryon) vertices of dimension $d$
(either a small momentum or meson mass).
This means that tree graphs start to contribute at order $q$
and $L$--loop graphs at order $q^{(2L+1)}$. Consequently, the low
energy constants appearing in ${\cal L}_{MB}^{(2)}$ are all finite.


\section{Generating functional to one loop}
\label{sec:genone}

In this section, we turn to the calculation of $S_M[j]+Z_{MB}[j,R_v]$ to
one loop, i.e. to order $q^3$ in the small momentum expansion. For doing
that,  we follow essentially the method used in ref.\cite{gss}. To be 
specific, one has to expand
\beq
{\cal L}_M^{(2)} +  {\cal L}_M^{(4)} - \bar{R}^a_v  \,
[A^{ab}_{(1)}]^{-1} \, R_v^b
\label{Zexp}
\eeq
in the functional integral Eq.(\ref{Zfinal}) around the classical
solution, $u_{\rm cl} = u_{\rm cl}[j]$, obtained by the variation
$\delta \, \int d^4x \,{\cal L}_{M}^{(2)} / \delta U$ to lowest order.
We need the first 
term in Eq.(\ref{Zexp}) to order $\xi^3$, the second one to order
$\xi$ and the last one to order $\xi^2$.
All other terms in
Eq.(\ref{ZMB}) start at order $q^2$ or higher and thus can not
contribute to the order $q^3$ one loop generating functional. The
calculation for the purely mesonic sector is identical to the one 
given in ref.\cite{gl85}. One thus arrives at a set of irreducible and
reducible diagrams as shown in Fig.~1. The sum of the reducible
diagrams is finite. Let us consider the
irreducible ones. As in the mesonic sector, we
chose the fluctuation variables $\xi$ in a symmetric form \cite{gl85},
\beq
\xi_R = u_{\rm cl} \, \exp\{i \, \xi /2\} \,\,\, , \quad 
\xi_L = u_{\rm cl}^\dagger \, \exp\{-i \, \xi /2\} \,\,\, , 
\eeq
with $\xi^\dagger = \xi$ traceless 3$\times$3 matrices. Consequently,
we have also
\beq
U =  u_{\rm cl} \, \exp\{i \, \xi \} \, u_{\rm cl} \,\,\, .
\eeq  
To second order in $\xi$, the covariant derivative $\nabla_\mu$, 
the chiral connection $\Gamma_\mu$ and the axial--vector $u_\mu$
take the form
\beqa
\Gamma_\mu &=& \Gamma_\mu^{\rm cl} + \frac{1}{4} \,[\, u_\mu^{\rm cl}, \xi
\, ] + \frac{1}{8} \, \xi 
\stackrel{\leftrightarrow}{\nabla}_\mu^{\rm cl}\, \xi + {\cal O}(\xi^3)
\nonumber \\ 
{} [ \nabla_\mu^{\rm cl} , \xi ] &=& \partial_\mu \, \xi + [ \, \Gamma_\mu^{\rm
  cl}, \xi \, ] \,\, ,\,\, \xi \stackrel{\leftrightarrow}{\nabla}_\mu^{\rm
  cl} \xi = \xi 
[ \nabla_\mu^{\rm cl} , \xi ] - [ \nabla_\mu^{\rm cl} , \xi ] \xi
\nonumber \\
u_\mu &=& u_\mu^{\rm cl} - [ \nabla_\mu^{\rm cl} , \xi ] + \frac{1}{8} \,
[\, \xi, [\, u_\mu^{\rm cl}, \xi\,]\,] + {\cal O}(\xi^3) \,\,\, .
\label{nablacl}
\eeqa
Notice that while $\nabla_\mu^{\rm cl}$ defined here acts on the
fluctuation variables (fields) $\xi (x)$, the covariant derivative
$\nabla_\mu$ defined in Eq.(\ref{nablaB}) acts on the baryon fields.
Inserting this into the expression for $A_{(1)}^{ab}$ and retaining
only the terms up to and including order $\xi^2$ gives
\beqa
A_{(1)}^{ab} &=& A_{(1)}^{ab, \, \rm cl}
+\frac{i}{4}\,<{\lambda^a}^\dagger \, [\,[ v \cdot u_{\rm cl}, \xi
\,], \lambda^b\,]> - D/F< {\lambda^a}^\dagger \, ( [S \cdot \nabla_{\rm
  cl} , \xi ] , \lambda^b \, )_\pm > \\
&+& \frac{i}{8}\,<{\lambda^a}^\dagger \, [\,\xi \, v \cdot 
\stackrel{\leftrightarrow}{\nabla}_{\rm cl} \xi ,
\, \lambda^b\,]> + \frac{1}{8} D/F< {\lambda^a}^\dagger \, ( 
\,[\,\xi, [\,S \cdot u_{\rm cl}, \xi \, ]\,], \lambda^b )_\pm >
+ {\cal O}(\xi^3) \,\,\, . \nonumber
\label{Aab1}
\eeqa
We have not made explicit in Eq.(\ref{Aab1}) the dependence 
of all operators on the external sources $j$ (see below).

We are now in the position to expand the fermion propagator $S$ to
quadratic order in the fluctuations making use of the relation
(for the moment, we suppress the flavor indices)
\beq
A_{(1)} \cdot S_{(1)} = {\bf 1} \quad .
\eeq
Following Gasser et al. \cite{gss}, we split $A_{(1)}$ into the free
and the interaction part,
\beq
A_{(1)} = d_0 + d_I = d_0 \, [1 + d_0^{-1} \, d_I] \quad .
\eeq
We tentatively assume the existence of the inverse of the free fermion
propagator. The interaction term $d_I$ admits a low energy expansion
starting at order $q$,
\beq
d_I = d_{I}^{cl} + d_{I}^1 + d_I^2 + \ldots \quad.
\eeq
Therefore, the fermion propagator to order $\xi^2$ reads
\beq
S_{(1)} = S_{(1)}^{\rm cl} - S_{(1)}^{\rm cl} \, d_1 \, S_{(1)}^{\rm cl} -
S_{(1)}^{\rm cl} \, d_2 \, S_{(1)}^{\rm cl} + S_{(1)}^{\rm cl} \, d_1 
\, S_{(1)}^{\rm cl} \, d_1 \, S_{(1)}^{\rm cl} + {\cal O}(\xi^3,q^3) \,\,\,. 
\label{Sexp}
\eeq
Here, $S_{(1)}^{\rm cl}$ denotes the full classical fermion
propagator, i.e. with all possible tree structures of the external
sources attached,
\beq
S_{(1)}^{\rm cl} = ( 1+ d_0^{-1} d_I^{\rm cl})^{-1} \, d_0^{-1} \,\,\,
, \eeq
where $d_I^{\rm cl}$ denotes the interactions with the Goldstone
bosons after expanding around the classical solution. The expanded
fermion propagator in Eq.(\ref{Sexp}) leads to an irreducible and a
reducible part in the generating functional. Let us first consider the
irreducible graphs. These are generated from the third and fourth term
in Eq.(\ref{Sexp}), respectively, and referred to as the tadpole and
the self--energy contribution, in order. Both are of ${\cal
  O}(\xi^2)$. The corresponding generating functional reads 
\beqa
Z_{\rm irr}[j,R_v^a] &=& \int d^4x \, d^4x' \, d^4y \, d^4y' \, 
\bar{R}_v^a (x) \, S_{(1)}^{bc, \,\rm cl} (x,y) \times \nonumber\\
& & \qquad\qquad\, \bigl[ \, \Sigma_2^{cd}(y,y) \,  \delta(y-y')
+ \Sigma_1^{cd} (y,y') \, \bigr] \, S_{(1)}^{de, \, \rm cl}(y',x')
\, R_v^e(x')
\label{Zirr}
\eeqa
in terms of the self--energy functionals $\Sigma_{1,2}$. These read
\beqa
\label{S12Vi}
\Sigma_1^{ab} &=& -\frac{2}{F_0^2} \, V_i^{ac} \, G_{ij} \, 
[A_{(1)}^{cd,cl}]^{-1} \, V_j^{db} 
=  -\frac{2}{F_0^2} \, V_i^{ac} \, G_{ij} \, 
S_{(1)}^{cd,cl} \, V_j^{db} \nonumber \\
\Sigma_2^{ab} &=& \frac{1}{8F_0^2} \biggl\{ D/F < {\lambda^a}^\dagger \, (
 \, [ \, \lambda^i_G \, , \, [ \, S \cdot u^{cl} , \lambda^j_G \,] \, ]\,  , 
\lambda^b \,)_\pm > \, G_{ij} \nonumber \\
& & \qquad \qquad \quad + i < {\lambda^a}^\dagger \, [ \,
\lambda^i_G (G_{ij} v \cdot \stackrel{\leftarrow}{d}_{jk} 
- v \cdot d_{ij} G_{jk} ) \lambda^k_G ,
\lambda^b \, ] > \biggr\} \nonumber \\
V_i^{ab} & = &  V_i^{ab(1)} + V_i^{ab(2)} \nonumber \\
V_i^{ab(1)} & = & \frac{i}{4\sqrt{2}} < {\lambda^a}^\dagger \, [\, [ v
\cdot u^{cl} , \lambda^i_G \,],  \, \lambda^b \, ] > \nonumber \\
V_i^{ab(2)} & = & - \frac{D/F}{\sqrt{2}}
< {\lambda^a}^\dagger \, ( \, \lambda^j_G \, S \cdot d_{ji}, \lambda^b
\, )_\pm > 
\eeqa
with $i,j,k= 1, \ldots ,8$ and $\lambda^i_G$ denote Gell--Mann's SU(3)
matrices, which are related to the ones in the physical basis by
$\lambda_p = (\lambda^4_G+i \lambda^5_G)/2$, $\lambda_n = (\lambda^6_G+
i \lambda^7_G)/2$, and so on.  
$G_{ij}$ is the full meson propagator \cite{gl85}
\beq
G_{ij} =  ( d_\mu \, d^\mu \, \delta^{ij} + \sigma^{ij} \, )^{-1}
\label{Mesprop}
\eeq
with
\beqa
[\nabla^\mu_{\rm cl}\, , \xi] &=& \frac{1}{\sqrt{2}} \lambda_G^j \,
d^\mu_{jk} \, \xi_k \,\,  \, , 
\quad \xi = \frac{1}{\sqrt{2}} \, \lambda_G^i \, \xi_i \,\, , \nonumber \\
d^\mu_{ij} &=& \delta_{ij} \, \partial^\mu + \gamma^\mu_{ij} \,\, ,
\stackrel{\leftarrow}{d}^\mu_{ij} = \delta_{ij}\,
\stackrel{\leftarrow}{\partial}^\mu - \gamma_{ij}^\mu \,\, , 
\nonumber \\
\gamma^\mu_{ij} &=& -\frac{1}{2} < \Gamma^\mu_{\rm cl} \, [ \, \lambda_G^i,
\lambda_G^j \, ] > \,\,\, , \nonumber \\
\sigma^{ij} &=& \frac{1}{8} <[ \, u^{\rm cl}_\mu , \lambda_G^i \, ][ \,
\lambda_G^j  , u_{\rm cl}^\mu   \, ] + \chi_+ \, \{ \, \lambda_G^i ,
\lambda_G^j \,\} > \,\,\, .
\label{MesHK}
\eeqa
Note that the differential operator $d_{ij}$ is related to the
covariant derivative $\nabla^\mu_{\rm cl}$ and it acts on the meson
propagator $G_{ij}$. 
The connection $\gamma_\mu$ defines a field strength tensor,
\beqa
\gamma_{\mu \nu} &=& \partial_\nu \, \gamma_\mu - \partial_\mu \,
\gamma_\nu + [ \gamma_\mu , \gamma_\nu ] \,\,\, , \nonumber \\
\, [d_\lambda \,, \, \gamma_{\mu \nu}]
 &=& \partial_\lambda \, \gamma_{\mu \nu}
+ [ \gamma_\lambda , \gamma_{\mu \nu} ] \,\,\, ,
\label{mesfs}
\eeqa
where we have omitted the flavor indices. For the calculations, it
helps to make use of the Bianchi--identity,
\beq
[d_\lambda \,, \, \gamma_{\mu \nu}] + [d_\nu \,, \, \gamma_{\lambda \mu}]
+ [d_\mu \,, \, \gamma_{\nu \lambda}] = 0 \,\,\, .
\eeq

\section{Heat kernel techniques}
\label{sec:heat}

In this section, we collect the heat kernel technique formulae which
are necessary to extract the divergent parts of the self--energy
functionals $\Sigma_{1,2} (x,y)$ in the coincidence limit $x \to y$.
This method is particularly useful since it maintains the underlying
symmetries. One considers the propagators in $d$--dimensional
Euclidean space. In the heat kernel representation, the divergences
appear as simple poles in $\epsilon = 4-d$ with residua that are
local polynomials of order $q^3$ in the fields. The latter can easily
be transformed back to Minkowski space.  
The reader familiar with these techniques might skip this section.
Detailed expositions of the method can be found in the reviews
\cite{gilkey}\cite{ball}.

Consider first an elliptic second--order differential operator of
the form (in $d$ Euclidean dimensions)
\beq
A = -d_\mu \, d_\mu + a(x) + \mu^2 \,\, , \quad d_\mu = \partial_\mu
+ \gamma_\mu \,\, ,
\label{esop}
\eeq
where $\gamma_\mu$ and $\sigma = a(x) + \mu^2$ are $C^\infty$ valued
matrix functions and $\gamma_\mu$ is anti--hermitian where as $A$ is 
hermitian. The heat kernel $G(t) = \exp \{ -A \,t \}$
satisfies the (diffusion) equation
\beq
\frac{\partial}{\partial \,t} G(t) + A \, G(t) = 0
\label{heateq}
\eeq
subject to the boundary condition $G(t=0) =1$. Splitting the heat
kernel into its free and interaction part,
\beq
G = G_0 \,  H \,\, , \quad A_0 = -\partial_\mu \, \partial_\mu  + \mu^2
\,\, , \eeq
the coordinate space representation of the free heat kernel $G_0$
reads
\beq
G_0 (x,y,t) = <x|G_0(t)|y> = <x|\exp\{-A_0\,t\}|y> 
= \frac{1}{(4\pi t)^{d/2}} \exp
  \biggl\{ -\mu^2\, t - \frac{z_\mu z_\mu}{4 t} \biggr\} \,\, ,
\eeq
with $z_\mu = x_\mu - y_\mu$.
$G_0$ satisfies the heat equation Eq.(\ref{heateq}) in terms of the
free operator $A_0$ and consequently we have for the interaction part
$H$
\beq
\biggr[ \frac{\partial}{\partial \,t} + (A-\mu^2) + \frac{1}{t}\,
z_\mu \, d_\mu \biggr] \, H(x,y,t) = 0 \,\,\, .
\label{Heq}
\eeq
Notice that the differential operator $d_\mu$ only acts on
$x$. The boundary condition in coordinate space reads $G(x,y,0)=
\delta^d (x-y)$. Eq.(\ref{Heq}) can be solved using the ansatz 
\beq
H(x,y,t) = \sum_{n=0}^\infty h_n(x,y) \, t^n
\eeq
which leads to the recurrence relation for the coefficient functions 
$h_n(x,y)$ (the so-called Seeley-deWitt coefficients)
\beqa
(n + z_\mu d_\mu)h_n(x,y) &=& -[ \, a(x) - d_\mu d_\mu ] \, h_{n-1} (x,y)
\quad (n \ge 1) \nonumber \\
z_\mu \, d_\mu \, h_0 (x,y) &=& 0 \quad .
\label{rec1}
\eeqa
The task is now to determine the first few coefficients in the
coincidence limit, $h_n| \equiv h_n (x,y)|_{z=0}$. It is easy to show
that
\beq 
d_\alpha \, z_\mu \, d_\mu \, h_n| = d_\alpha \, h_n| \,\,\, .
\eeq
Similarly, for a string of $m$ differential operators $d_\alpha
d_\beta \ldots d_\omega$, one finds
\beqa
d_\alpha \ldots d_\omega \, h_n(x,y)|_{z=0} &=& -\frac{1}{m+n} 
\biggl\{ d_\alpha
\ldots d_\omega \, (a - d_\mu \, d_\mu)h_{n-1} (x,y) + P_{\alpha \ldots
  \omega} \, h_n (x,y) \biggr\} \biggr|_{z=0} \,\, , \nonumber \\
P_{\alpha \ldots \omega} &=& d_\alpha \ldots d_\omega \, z_\mu d_\mu -
m \,  d_\alpha \ldots d_\omega \,\,\, .
\eeqa
The first three heat coefficients $h_{0,1,2}$ follow as
\beq
h_0| = 1 \,\, , \, \,  h_1| = - a \, \, , \, \, 
h_2| = \frac{1}{2}\, a^2 - \frac{1}{6} \,[d_\mu,[d_\mu,a]]+
\frac{1}{12} \, (\gamma_{\mu \nu})^2 \,\, ,
\label{coeffs}
\eeq
with 
\beq
\gamma_{\mu \nu} = \partial_\mu \, \gamma_\nu -\partial_\nu  \,\gamma_\mu + 
[ \, \gamma_\mu , \gamma_\nu \,] = [\, d_\mu \, , \,d_\nu\,] \,\,\, . 
\eeq
The first few derivatives in the coincidence limit read
\beqa
d_\mu \, d_\nu \, h_0|  &=& \frac{1}{2}\,\gamma_{\mu \nu} \,\, , \, \,
d_\lambda \, d_\mu \, d_\nu \, h_0|  = \frac{1}{3}\, \biggl\{ [ \,
d_\lambda , \gamma_{\mu \nu}\,] + [\, d_\mu , \gamma_{\lambda \nu} \,]
\biggr\}\,\, , \nonumber \\
d_\alpha \, d_\alpha \, d_\mu \, d_\mu \, h_0| &=& \frac{1}{2} \,
\gamma_{\mu \alpha} \, \gamma_{\mu \alpha} \,\, ,\, \,
d_\mu \, h_1| = -\frac{1}{2}\, [\, d_\mu ,a \,] + \frac{1}{6} \, [
\, d_\nu , \gamma_{\mu \nu}\,] \,\, , \nonumber \\ 
d_\mu \, d_\mu \, h_1| &=& -\frac{1}{3}\, [\, d_\mu ,[\, d_\mu ,a \,]]
 + \frac{1}{6} \, \gamma_{\mu \alpha} \, \gamma_{\mu \alpha} \,\, . 
\label{dercoeff}
\eeqa 
We also need the differential operator acting from the right. It is
defined in complete analogy to Eq.(\ref{Heq}) via
\beq
\biggr[ \frac{\partial}{\partial \,t} + (A-\mu^2) + \frac{1}{t}\,
z_\mu \, d_\mu^x \biggr] \, ( H   \, \stackrel{\leftarrow}{d}_\nu^y )
=  \frac{1}{t} \, {d}_\nu^x \, H\,\,\, ,
\label{Heqleft}
\eeq
and the pertinent recurrence relations read
\beqa
(n + z_\mu d_\mu) \biggl( h_n(x,y) \stackrel{\leftarrow}{d}_\nu^y \biggr)
 &=& d_\nu \, h_n(x,y) 
- (A-\mu^2) \, h_{n-1} (x,y) \,\stackrel{\leftarrow}{d}_\nu^y
\quad (n \ge 1) \nonumber \\
z_\mu \,( d_\mu \, h_0 (x,y) \, \stackrel{\leftarrow}{d}_\nu^y )
 &=& d_\nu \, h_0(x,y) \quad ,
\label{rec1left}
\eeqa
and in the coincidence limit we have for the first derivatives
\beqa
d_\mu \, h_0 \, \stackrel{\leftarrow}{d}_\nu| &=& d_\mu\,d_\nu\,h_0|
\,\, , \quad \{ d_\lambda,d_\mu\}\, h_0 \,
\stackrel{\leftarrow}{d}_\nu| = d_\lambda \,d_\mu\, d_\nu\, h_0| \,\, ,
\nonumber \\
h_0 \, \stackrel{\leftarrow}{d}_\nu \, \stackrel{\leftarrow}{d}_\mu|
&=& d_\nu \, d_\mu\, h_0 |
\,\, , \quad d_\mu \, h_0 \, \stackrel{\leftarrow}{d}_\nu 
\, \stackrel{\leftarrow}{d}_\lambda| = \frac{1}{6} \, ( \, [ 
d_\lambda , \gamma_{\mu \nu}] + [d_\nu\,, \gamma_{\mu \lambda}]) \, \,\, ,
\nonumber \\
h_1 \, \stackrel{\leftarrow}{d}_\mu| &=& d_\mu \, h_1| - \frac{1}{3}
[d_\nu , \gamma_{\mu \nu}] = -\frac{1}{2}[d_\mu , a] - \frac{1}{6}
[d_\nu , \gamma_{\mu \nu}] \,\, .
\eeqa
We can now construct the propagator corresponding to Eq.(\ref{Mesprop}),
\beq
G(x,y) = \int_0^\infty dt \, G(x,y,t) \,\,\, ,
\eeq
which is a well--defined operator  for all $x \ne y$. At large
distances the behavior of $G(x,y)$ will be controlled by the form of
the interaction part at large $t$. One has a well defined behavior in
the infrared region, since the meson propagator in
Eq.(\ref{Mesprop}) is nothing less than the free propagator. 
The short--distance
(UV) behavior will be controlled by the form of the interaction part
$H$ at small $t$. $G(x,y)$ admits an asymptotic expansion for $t \to 0$
in terms of the Seeley--deWitt coefficients,
\beqa
\label{Gxy}
G(x,y) &=& \sum_{n=0}^\infty \, G_n(x,y) \, h_n (x,y)
\nonumber \\
G_n(x,y) &=& \int_0^\infty dt \, (4\pi t)^{-d/2} \, \exp \{ |x-y|^2 / 4t
\} \, t^n \nonumber \\ &=& \frac{1}{(4\pi)^{d/2}} \biggl[
 \frac{4}{|x-y|^2} \biggr]^{\frac{d}{2}-n-1}\, \Gamma (\frac{d}{2}-n-1) \,\, .
\eeqa
Observe that $H$ tends to 1 as $t \to 0$ so that the proper--time 
integration fails to converge at the lower limit when $x=y$ for $d\ge
2$. This is reflected in the singular behavior of $G(x,y)$ as $|x-y|
\to 0$. In particular,
the functions $G_n$ contain divergences for $d=4$. Using an
$\overline{MS}$ subtraction scheme, one rewrites the expansion for $G(x,y)$
as \cite{jacko}
\beq
G(x,y) = G_0(x,y) \,  h_0(x,y) + \sum_{n=1}^{2} \, R_n(x,y) \, h_n (x,y)
+ \bar{G} (x,y)
\eeq
with
\beqa
\label{G0R12}
G_0(x) &=& \frac{1}{4\pi^{d/2}} \, \Gamma \biggl( \frac{d}{2}-1
\biggr) \, |x|^{2-d}
\to  \frac{1}{4\pi^2 |x|^2} \quad {\rm for} \, \,\, d \to 4 \, \, 
\nonumber \\
R_1(x) &=& 
\frac{1}{16\pi^2} \biggl[
\frac{2}{\epsilon}\mu^{-\epsilon} + \frac{\Gamma(d/2-2)}{\pi^{d/2-2}}
\, |x|^{4-d} \biggr]
\to -\frac{1}{16\pi^2} \, \bigl\{
\gamma + \ln \pi + \ln(\mu^2 \, |x|^2) \bigr\} \,\, , \label{R12} \\ 
R_2(x) &=& 
\frac{-1}{32\pi^2} \biggl\{
\frac{|x|^2}{\epsilon}\mu^{-\epsilon} - \frac{\Gamma(d/2-3)}{2\pi^{d/2-2}}
\, |x|^{6-d} \biggr\}
\to \frac{1}{64\pi^2} \, |x|^2 \,\biggl[
\gamma + \ln \pi -1 + \ln(\mu^2 \, |x|^2) \biggr] \,\, , \nonumber
\eeqa
in the coincidence limit $x=y$. 
Here, $\mu$ is a mass scale introduced on dimensional grounds and
$\gamma= 0.5772$ is the Euler--Mascheroni constant. Notice that $R_1$
and $R_2$ are regular as $\epsilon \to 0$ and $\bar G$ has no poles in
$\epsilon$ and is regular for $x=y$, even for two derivatives. 
$G(x,y)$ is independent of the
scale $\mu$ and any arbitrariness due to the regularization scheme is
compensated by $\bar G (x,y)$. By analytic continuation in $d$
from $d < 2$ it follows that
\beq 
G(x,x) = \frac{2}{\epsilon} \frac{\mu^{-\epsilon}}{16 \pi^2} \, h_1
(x,x) + {\rm finite} =
\frac{2}{\epsilon} \frac{\mu^{-\epsilon}}{16 \pi^2} \, h_1
(x,x) + \bar{G}(x,x) \,\,\, .
\label{GHK}
\eeq
Similarly, we need find for the first derivative of $G(x,y)$,
\beq
d_\mu \, G(x,y) = d_\mu \, \sum_{n=0}^\infty G_n \, h_n = 
\sum_{n=0}^\infty (\partial_\mu \, G_n) \, h_n +
\sum_{n=0}^\infty G_n \, (d_\mu \,h_n ) \,\, \, .
\eeq
In the coincidence limit, this simplifies to
\beq 
d_\mu \, G(x,x) = \frac{2}{\epsilon} \frac{\mu^{-\epsilon}}{16 \pi^2}
\,  d_\mu \, h_1 (x,x) + {\rm finite} =
\frac{2}{\epsilon} \frac{\mu^{-\epsilon}}{16 \pi^2} \, d_\mu \, h_1
(x,x) + d_\mu \, \bar{G}(x,x) \,\,\, ,
\eeq
which means that all divergences are encoded in the coefficient $h_1
(x,x) = h_1|$ and its first derivative, $d_\mu \, h_1|$. 
This technology is sufficient to determine the leading
divergences in the meson sector and the ones related to the tadpole
graph.

In case of the self--energy graph, we need a modification as proposed
in ref.\cite{ecker}. Consider the following differential operator,
\beq
\Delta  = - (v \cdot d)^2 + a(x) + \mu^2 \,\, , \quad d_\mu = 
\partial_\mu + \gamma_\mu \,\,\, .
\eeq
The heat kernel $J(t) = \exp\{-\Delta t\}= J_0 \, K$ 
can be split again into its free part,
\beq
\label{J0}
J_0 (t) =  \frac{1}{\sqrt{4\pi t}} \exp
  \biggl\{ -\mu^2\, t - \frac{[v \cdot (x-y)]^2}{4 t} \biggr\} \,\, , 
\eeq
and the interaction part $K$, which satisfies the equation,
\beq
\biggr[ \frac{\partial}{\partial \,t} - (v \cdot d)^2+ a(x) 
+ \frac{1}{t}\, v \cdot (x-y) \, v \cdot d \, \biggr] \, K(x,y,t)
= 0 \,\,\, ,
\label{H2eq}
\eeq
using $(v \cdot v) =1$. It is important to stress the difference to the
previous case. Because $v \cdot d$ is a scalar, one has
essentially reduced the problem to a one--dimensional one, i.e.
$v \cdot d$ is a one--dimensional differential operator in the
direction of $v$. Because of this,
it is advantageous to modify the heat kernel expansion for $K$,
\beq
K(x,y,t) =  g(x,y) \, \sum_{n=0}^\infty k_n(x,y) \, t^n \,\, ,
\eeq
where the function $g(x,y)$ is introduced so that one can fulfill the
boundary condition  in the coordinate--space representation
(see below). The explicit form of the function $g$ is not needed to
derive the recurrence relations for the heat kernel. It is given later
when products of singular operators are constructed (see appendix~B).
The pertinent recurrence relations are
\beqa
[ \, n + v \cdot (x-y) \, v \cdot d \, ] \, g(x,y) \,
k_n(x,y) &=& -[ \, a(x) - (v \cdot d )^2 \, ] \, g(x,y) \, k_{n-1} (x,y)
\quad (n \ge 1) \nonumber \\
{}[ \, v \cdot (x-y) \, v \cdot d  \, ] \,
 g(x,y) \, k_0 (x,y) &=& 0 \quad .
\label{rec2}
\eeqa
This can be cast into the form
\beqa
(v \cdot d)^m \, g(x,y) \, k_n (x,y) &=& \frac{-1}{m+n} \biggl\{
(v \cdot d)^m [ -(v \cdot d)^2 + a(x) ] g(x,y) \, k_{n-1}
(x,y) \nonumber \\ & & \qquad \qquad \qquad
+ P^m \, g(x,y) \, k_n (x,y) \biggr\} \nonumber \\
P^m &=& (v \cdot d)^m (v \cdot z) (v \cdot d) - m \, 
(v \cdot d)^m \,\,\, .
\eeqa
It is easy to show that $P^m \, g \, k$ vanishes in  the coincidence
limit for all $m \ge 0$. Demanding now
\beq
v \cdot \partial \, g(x,y) = 0 \,\,\, ,
\eeq
the recurrence relation takes the particular simple form
\beq
(v \cdot d)^m \, k_n (x,y)|_{z=0}  = \frac{-1}{m+n} \biggl\{
(v \cdot d)^m [ -(v \cdot d)^2 + a(x) ]  \, k_{n-1} (x,y)
\biggr\} \biggr|_{z=0} \,\,\, .
\eeq
It is straightforward to read off the lowest Seeley--deWitt
coefficients and their derivatives,
\beqa
k_0| &=& 1 \,\, , \, \,  (v \cdot d)^m \, k_0|=0 \,\, ,\,\,
(v \cdot d)^m \, k_0 \, (v \cdot
\stackrel{\leftarrow}{d})^n| = 0 \, \, ,\,\,
k_1|=-a\,\, , \nonumber \\
(v \cdot d) \, k_1| &=&-\frac{1}{2} \, [ \, v \cdot d , a \, ]
\,\, , \, \, 
(v \cdot d)^2 \, k_1|=-\frac{1}{3} \, [ \, v \cdot d , 
\, [ \, v \cdot d ,a \, ]] \,\, , \nonumber \\
k_2| &=&\frac{1}{2} \, a^2 - \frac{1}{6} \, [ \, v \cdot d, [ \, 
v \cdot d , a \, ] \, ] \,\, , \, \, k_1 \, v \cdot
\stackrel{\leftarrow}{d}| = v \cdot d \, k_1| \,\,\, .
\eeqa
The propagator is given as the integral
\beqa
\label{Jn}
J(x,y) &=& \Delta^{-1}(x,y) = \int_0^\infty dt \, J(x,y,t) \nonumber
\\ J(x,y) &=& \sum_{n=0}^\infty J_n(x,y) \, k_n (x,y) \nonumber \\
J_n (x,y) &=&  g(x,y)  \, 
\int_0^\infty  \frac{dt}{\sqrt{4\pi t}} \exp
  \biggl\{ -\mu^2\, t - \frac{[v \cdot (x-y)]^2}{4 t} \biggr\} \, 
 t^n  \,\,\, .
\eeqa
Since the particle propagator $A^{ab}_{(1)}$ is massless, one must
keep $\mu^2 \ne 0 $ in intermediate steps to get a well defined heat
kernal representation without infrared singularities.
For later use, we also need the operator $v \cdot d$ acting on
$J(x,y)$,
\beq
v \cdot d \, J(x,y) = \sum_n(v \cdot \partial \, J_n (x,y) \, ) \, k_n
(x,y) + \sum_n J_n(x,y) v \cdot d \, k_n(x,y) \,\,\, .
\eeq
We are now in the position to apply these methods to the problem under
investigation.

\section{Renormalization of the tadpole graph}
\label{sec:tadpole}

In this section, we consider the renormalization of the tadpole
contribution $\Sigma_2^{ab} (y,y)$. This is done in Euclidean space
letting $x^0 \to -i \, x^0$, $v^0 \to i \, v^0$, $v \cdot \partial \to 
-v \cdot \partial$, $S^0 \to -i \, S^0$ and $S \cdot u \to  S \cdot u$.
To show how this calculation works,
we split the tadpole into  two terms as in Eq.(\ref{S12Vi}), 
$\Sigma_2^{ab} = \Sigma_2^{ab(1)} + \Sigma_2^{ab(2)}$, with the 
first term being proportional to the meson progagator $G_{ij}$. 
The meson propagator is an operator of the type Eq.(\ref{esop}) and
using the result of Eq.(\ref{GHK}), we can identify
\beq
h_1^{ij}| = - \sigma^{ij} \,\,\, ,
\eeq
with $\sigma^{ij}$ given in Eq.(\ref{MesHK}). The divergent part  
of the tadpole can be cast in the form (rotated back to Minkowski space) 
\beq
\Sigma_2^{ab , {\rm div}} (y,y) = \frac{1}{(4 \pi F_0 )^2}  
\frac{2}{\epsilon} \, \, \hat{\Sigma}_2^{ab} (y,y) \,\,\, ,
\label{rentad}
\eeq
with  $\hat{\Sigma}_2^{ab} (y,y)$  a finite monomial in the
fields of chiral dimension three. As an example, let us take a closer
look at $\Sigma_2^{ab(1)}$. For that, consider the object
\beq
I^{ab} = <{\lambda^a}^\dagger \, ( \, [ \lambda_G^i , [\, A , \lambda_G^j \,]]
, \lambda^b )_\pm > \, <B^{ij}>
\eeq
with $A= S \cdot u$ and $B^{ij} = \chi_+ \, \{\lambda_G^i, \lambda_G^j\}$
for a typical case. Using the completeness and various trace relations
(collected in appendix~A), this can be cast into the form
\beqa
I^{ab} &=& 8 \, < ( \, {\lambda^a}^\dagger , \lambda^b \,)_\pm > \, 
<A \, B> + 8 \, <  {\lambda^a}^\dagger (\, B , \lambda^b \,)_\pm > \, <A>
\nonumber \\
&-& 8 \, <  {\lambda^a}^\dagger (\, A , \lambda^b \,)_\pm > \, <B>
- 12 \, <  {\lambda^a}^\dagger (\, \{ \,A , B\, \} , \lambda^b \,)_\pm >
\,\,\, . \label{exAB}
\eeqa
Using now the trace relations
\beq
< u_\mu  > = 0 \,\, , \quad < S \cdot u > = 0 \,\,\, ,
\eeq
we find
\beqa
\hat{\Sigma}_2^{ab(1)} (y,y) &=& -\frac{1}{8}\, D/F \, \biggl\{
-\frac{3}{2} < {\lambda^a}^\dagger  \, 
( \{ S \cdot u, u \cdot u + \chi_+ \} , \lambda^b)_\pm > \nonumber \\
&-&  < {\lambda^a}^\dagger  \, ( S \cdot u, \lambda^b )_\pm><u \cdot
u + \chi_+>  - 2\,< {\lambda^a}^\dagger  \, ( u_\mu, \lambda^b )_\pm>
\, <u^\mu \, S \cdot u > \biggr\} \nonumber \\ 
&-& \frac{1}{4}D \, < \, {\lambda^a}^\dagger \,\lambda^b \, > < S \cdot u \,
(\, u \cdot u + \chi_+) \, >   \, \,\,\, . 
\label{S2div1} 
\eeqa 
We turn to the divergence structure of $\Sigma_2^{ab(2)} (y,y)$. For
that, we need the following heat kernel expansion (omitting all traces
and SU(3) matrices)
\beq
G_{ij} ( v \cdot \stackrel{\leftarrow}{d}^y )_{jk} - (v \cdot 
d^x)_{ij} G_{jk} \stackrel{x \to y}{\rightarrow} \sum_n G_n [(h_n)_{ij} v \cdot
\stackrel{\leftarrow}{d}^y_{jk}] - G_n [v \cdot d_{ij} (h_n)_{jk}]
\,\, .
\eeq
Using now Eq.({\ref{dercoeff}), Eq.(\ref{R12}) and Eq.(\ref{mesfs}), 
this turns into
\beq
\frac{2}{\epsilon}\frac{\mu^{-\epsilon}}{16\pi^2} \left[ (h_1)_{ij} \,
v  \cdot \stackrel{\leftarrow}{d}_{jk}^y - v \cdot d_{ij} \, (h_1)_{jk}
\right] = \frac{2}{\epsilon}\frac{\mu^{-\epsilon}}{16\pi^2} \left[
-\frac{1}{3} \, v^\mu \, [d_\nu , \gamma_{\mu \nu}]_{ik} \, \right] \,\,
.
\eeq
Reinstating the prefactors and SU(3) matrices, we find after some
lengthy but straightforward algebra for $\hat{\Sigma}^{ab(2)}_2 (y,y)$
\beq
\hat{\Sigma}^{ab(2)}_2 (y,y) = -i \frac{1}{4} \, < {\lambda^a}^\dagger \, [
[\nabla^\mu \, ,  \Gamma_{\mu \nu} \, v^\nu \,] , \, \lambda^b \,] > 
\,\,\, ,
\label{S2div2}
\eeq
with $\Gamma_{\mu \nu}$ the standard field strength tensor related to 
$\Gamma_\mu$ and we dropped the indices 'cl', see Eqs.(\ref{MesHK}).
The divergences encoded in the tadpole, cf.~Eq.(\ref{rentad}), are 
therefore determined, and the finite polynomial of order $q^3$ reads
\beq
\hat{\Sigma}_2^{ab} (y,y) = \hat{\Sigma}_2^{ab(1)} (y,y) + 
\hat{\Sigma}_2^{ab(2)} (y,y) \quad ,
\eeq
as given in  Eqs.({\ref{S2div1},\ref{S2div2}).

\section{Renormalization of the self--energy graph}
\label{sec:selfenergy}

In this section, we consider the renormalization of the self--energy
contribution $\Sigma_1^{ab} (y,y)$. The divergences are due to the
singular behavior of the product of the meson and the baryon
propagators
\beq
 G_{ij} (x,y) \, [A_{(1)}^{ab}]^{-1} (x,y)
\eeq
in the coincidence limit $x \to y$. This expression is directly 
proportional to the full
classical fermion propagator $S_{(1)}^{ab, {\rm cl}}$ as discussed in
section~\ref{sec:genone}.\footnote{From now on, we drop the index
  'cl'.} However, this differential operator is not
elliptic and thus not directly amenable to the heat--kernel
expansion. Consider therefore the object \cite{ecker}
\beq
S_{(1)}^{ab} = i \, [i \, A_{(1)}^{ac}]^\dagger \, [ (i \,
A_{(1)}^{cd} ) (i \, A_{(1)}^{cd} )^\dagger ]^{-1} \,\,\, .
\label{D}
\eeq
In fact, the operator in the square brackets in Eq.(\ref{D}) is
positive definite and hermitian. Furthermore, it is a one--dimensional
operator in the direction of $v$ and we can use the heat kernel
methods spelled out in section~\ref{sec:heat} for such type of
operators. In Euclidean space, we have
\beqa
i \, A_{(1)}^{ab} &=& -<{\lambda^a}^\dagger \, [v \cdot \nabla \,,
\lambda^b]> - i D/F < {\lambda^a}^\dagger \, ( S \cdot u,
\lambda^b\,)_\pm > \nonumber \\
(i \, A_{(1)}^{ab})^\dagger &=& +<{\lambda^a}^\dagger \, [v \cdot \nabla \,,
\lambda^b] > - i D/F < {\lambda^a}^\dagger \, ( S \cdot u,
\lambda^b\,)_\pm > \, \, \, ,
\eeqa
or in a more convenient form for later use
\beqa
i \, A_{(1)}^{ab} &=& -v \cdot d^{ab} - i D/F < {\lambda^a}^\dagger 
\, ( S \cdot u, \lambda^b\,)_\pm > \equiv -v \cdot d^{ab} + \rho^{ab} 
\nonumber \\
v \cdot d^{ab} &=&\delta^{ab} \, v \cdot \partial + v \cdot \gamma^{ab}
\,\, , \, \, v \cdot \gamma^{ab} = < {\lambda^a}^\dagger [ v \cdot \Gamma , 
\lambda^b]> \,\, .
\eeqa
We note that it is important that $\rho^{ab}$ does not contain any 
differential operator. After some algebra one gets
\beq
( i \, A_{(1)}^{ac} ) (i \, A_{(1)}^{cb} )^\dagger = 
- v \cdot d^{ac} \, v \cdot d^{cb} + a^{ab}
\eeq
with 
\beqa
a^{ab} &=&a_1^{ab} + a_2^{ab} \,\,\, , \nonumber \\ 
a_1^{ab}&=& i  D/F \, <{\lambda^a}^\dagger (\, [v \cdot \nabla , S \cdot u ]
,\, \lambda^b)_\pm  >  \\
a_2^{ab}&=&- <{\lambda^a}^\dagger ( D/F \, S \cdot u , (
D/F \, S \cdot u , \lambda^b )_\pm )_\pm > 
 + {4 \over 3} D^2 \, <{\lambda^a}^\dagger \, S \cdot u>
< S \cdot u \, \lambda^b> \,\,\, \nonumber .
\eeqa
The last term in this equation is typical for SU(3), i.e. such a type of
term does not appear in the analogous SU(2) calculation
\cite{ecker}. It can be traced back to the use of the completeness
relation, Eq.(\ref{comprel}), and that fact that $< [\lambda^a \, , \,
O ]>=0$ for $O = O^b \lambda^b \in$ SU(3),
\beqa
& & <{\lambda^a}^\dagger \, (O, \lambda^c)_\pm> \, <{\lambda^c}^\dagger \,
(O', \lambda^b)_\pm >^\dagger \nonumber \\
&=& <{\lambda^a}^\dagger \, (O, ({O'}^\dagger,\lambda^b)_\pm)_\pm > -
\frac{D^2}{3} < \{ {\lambda^a}^\dagger , O \} > \, <
\{ {O'}^\dagger,\lambda^b \} > \,\,\, .
\eeqa
The corresponding singularities of $\Sigma_1^{ab}$ can be extracted
in terms of singular products of $G_{0,1} (x)$ and $J_0(x)$ as listed
in detail in Eqs.(44-51) in Ecker's paper \cite{ecker}. In appendix B,
we give these for completeness and show how one derives them (for one 
particular example). The heat kernel expansion of $G_{ij} (x,y) \, 
[A_{(1)}^{ab}]^{-1} (x,y)$ takes the form
\beqa
X^{ij,cd} &=& i \, \biggl\{ G_n  h_n^{ij} (v \cdot d^{cd} + \rho^{cd})
J_m k_m \biggr\} \nonumber \\
&=& i \, \biggl\{ ( G_n v \cdot \partial \, J_m ) \, h_n^{ij}\delta^{cd}
k_m + G_n J_m h_n^{ij} (v \cdot d^{cd} + \rho^{cd}) k_m \biggr\} \, .
\label{Xij}
\eeqa
This operator has to be sandwiched between the different interactions
$V_i^{ab(1,2)}$ defined in Eq.(\ref{S12Vi}) and one evaluates the
corresponding products. This procedure is very general, i.e. it holds
for all interactions of a structure like given in Eq.(\ref{Xij}). In 
particular, the two-- and three--flavor case can be treated on the same
footing. We have to differentiate between three types of products,
\beqa
1) && V_i^{ac(1)} (x) \, X^{ij,cd} (x,y) \, V_j^{db(1)} (y) \nonumber \\
2) && V_i^{ac(1)} (x) \, X^{ij,cd} (x,y) \, V_j^{db(2)} (y) 
+ (1 \leftrightarrow 2) \nonumber \\
3) && V_i^{ac(2)} (x) \, X^{ij,cd} (x,y) \, V_j^{db(2)} (y) \,\, .
\eeqa
Notice that the interactions $V_i$ depend on different space--time
points. Consider case 1). It is straightforward to show (using
Eqs.(44)-(51) of ref.\cite{ecker}) that the divergent part of this
operator takes the form
\beqa
\label{coveq}
\frac{i}{(4\pi )^2} \, \frac{2}{\epsilon} \, V_i^{ac} (x)
&& \!\!\!\!\!\!\!
\biggl\{  -2 v \cdot \partial \delta(x-y) \, h_0^{ij} (x,y)
\delta^{cd} k_0 (x,y) \nonumber \\
&& \quad + 2 \delta(x-y) \, h_0^{ij} (x,y) (v \cdot d^{cd} + \rho^{cd})
k_0 (x,y) \biggr\} V_j^{db} (y) \,\, .
\eeqa
After partial integration, one can perform the coincidence limit and
is left with the two operators $O^{ab}_{1,2}$ given below. In the
second and third case, at least one of the interactions contains a 
spin--operator $S_\mu$ and one covariant derivative acting on the
meson propagator. It is convenient to rewrite the interaction in a way
that it only contains flavor matrices,
\beq
V_i^{ac(2)} (x) \equiv V_k^{ac(2)} \, S_\mu \, d_\mu^{ki} \,\, .
\eeq
In the coincidence limit, we get for case 2):
\beqa
\frac{i}{(4\pi )^2} \, \frac{2}{\epsilon} \, \delta (x-y) && 
\!\!\!\!\!\!\!
\biggl[ -2 \biggl\{ V_i^{ac(1)} (x) \, S \cdot \stackrel{\leftarrow}{d} \,
h_0\,\, d \cdot v \bigr|^{ij} \, V_j^{cb(2)} (x) \nonumber \\ 
&& \quad + V_i^{ac(2)} (x) \, v \cdot \stackrel{\leftarrow}{d} \,
h_0 \, \, d \cdot S \bigr|^{ij} \, V_j^{cb(1)} (x) \, \biggr\} \biggr] \,\, .
\eeqa
This leads to the operator $O_3^{ab}$. Case 3) is more involved. We
find
\beqa
i \, V_i^{ac(2)} (x) \, S_\mu
&&  \!\!\!\!
\biggl\{ \partial_{x \mu} G_n (x,y) \stackrel{\leftarrow}{\partial}_{y\nu} \,
\, v \cdot \partial \, J_m (x,y) \, h_n^{ij} (x,y) \, k_m^{cd} (x,y) 
\nonumber \\
&&  \!\!\!\!
+ G_n (x,y) \, \, v \cdot \partial \, J_m (x,y) \, (d_\mu h_n (x,y) 
\stackrel{\leftarrow}{d}_\nu)^{ij} \, k_m^{cd} (x,y) 
 \\
&&  \!\!\!\!
+  \partial_{x \mu} G_n (x,y) \stackrel{\leftarrow}{\partial}_{y\nu} \,
 \, J_m (x,y) \, h_n^{ij} (x,y) \,((v \cdot d + \rho) k_m (x,y))^{cd} 
\nonumber \\
&&  \!\!\!\!
+ G_n (x,y) \,  \, J_m (x,y) \, (d_\mu h_n (x,y) 
\stackrel{\leftarrow}{d}_\nu)^{ij} \, ((v \cdot d + \rho) k_m
(x,y))^{cd} \biggr\} \, S_\nu \, V_j^{cd(2)} (y) \,\, .
\nonumber
\eeqa
Notice that terms linear in $\partial_\mu G_n$ vanish in the
coincidence limit. In what follows, we need the terms with $n=m=0$ or
$n=0,m=1$ or $n=1,m=0$,
whereas in the previous cases only $m=n=0$ was relevant.
Case 3) thus generates much more terms and leads to the operators
$O^{ab}_i$, $i=4, \ldots , 16$, as listed below. We remark that the
divergences are given by the singular products $G_n \, J_m$ and
derivatives thereof. This destroys covariance since only partial
derivatives are involved. However, by appropriately combining terms
one can restore covariance \cite{ecker}. The $O_i^{ab}$ given below
are such combinations. This restoration of covariance serves as an
important check on the calculation.
Rotating back to Minkowski space, one finds after
some lengthy algebra a local  functional $\Sigma_1^{ab} (y)$,
\beq
\Sigma_1^{ab , {\rm div}} (y,y) = \frac{1}{(4 \pi F_0 )^2}  
\frac{2}{\epsilon} \, \, \delta^4 (x-y) \, \sum_{i=1}^{16}\,\,
 \hat{\Sigma}_{1,i}^{ab} (y) \, \,\, ,
\label{S1local}
\eeq
where we have decomposed the lengthy expression for $\Sigma_1^{ab , 
{\rm div}} (y,y)$ in such a way that one can most easily recover the
SU(2) result, compare Eq.(53) of ref.\cite{ecker}. In appendix~C, we
list the  operators corresponding to the three cases discussed above
and the resulting contributions to the divergent part
of the self--energy functional.

\section{The counterterm Lagrangian}
\label{sec:LMB3}

We are now in the position to enumerate the full counterterm
Lagrangian at order $q^3$. To bring it in a more compact form, 
powers of the spin matrix are reduced with the help of
Eq.(\ref{spinrel}). We also use  the curvature relation
\beq
\Gamma_{\mu \nu} = \frac{1}{4}[u_\mu, u_\nu]  - \frac{i}{2} F^+_{\mu
  \nu} \,\, .
\eeq
To separate the finite parts in dimensional regularization, we follow
the conventions of \cite{gl85} to decompose the irreducible one--loop
functional into a finite and a divergent part. Both depend on the
scale $\mu$:
\beqa
\label{split}
\delta^4(x-y) \, \Sigma^{ab}_2 (x,y) + \Sigma^{ab}_1 (x,y) & = &
\delta^4(x-y) \, \Sigma^{ab, {\rm fin}}_2 (x,y, \mu) 
+ \Sigma^{ab, {\rm fin}}_1 (x,y, \mu) \nonumber \\
&& -\frac{2 L(\mu )}{F_0^2} \, \delta^4 (x-y) \, [ \hat{\Sigma}_2^{ab} (y) +
\hat{\Sigma}_1^{ab} (y) \, ] \,\,\, ,
\eeqa
with
\beq
L(\mu ) = \frac{\mu^{d-4}}{(4 \pi)^2} \biggl\{ \frac{1}{d-4} -
\frac{1}{2} [ \log(4\pi) + 1 - \gamma  ] \biggr\} \,\, .
\eeq
The generating functional can then be renormalized by introducing the
counterterm Lagrangian
\beq \label{LCT}
{\cal L}_{MB}^{(3)\, {\rm ct}} (x) = \frac{1}{(4 \pi F_0 )^2} \, \sum_i
d_i \, \bar{H}^{ab}_v (x) \, \tilde{O}^{bc}_i (x) \, H^{ca}_v (x)
\eeq
where the $d_i$ are dimensionless coupling constants and the field
monomials $\tilde{O}^{bc}_i (x)$ are of order $q^3$. The low--energy 
constants $d_i$ are decomposed in analogy to Eq.(\ref{split}),
\beq
d_i = d_i^r (\mu ) + (4 \pi)^2 \, \beta_i \, L(\mu ) \,\,\, .
\eeq
The $\beta_i$ are dimensionless functions of $F$ and $D$ constructed
such that they cancel the divergences of the one--loop functional. 
They are listed in table~1 together with the corresponding operators
$\tilde{O}^{bc}_i (x)$. Notice that in this table the quantity $\chi_-
= u^\dagger \chi u^\dagger - u \chi^\dagger u$ never appears because
of the relation
\beq
[\nabla^\mu , u_\mu] = \frac{i}{2}\chi_- - \frac{i}{4} <\chi_->  \,\,\, .
\eeq
The operators listed in table~1 constitute a complete set for the 
renormalization of the irreducible tadpole and self--energy functional
for {\it off-shell} baryons. These are the terms where the covariant 
derivative acts on the baryon fields. As long as one is only
interested in Green functions with {\it on-shell} baryons, the number
of terms can be reduced considerably by invoking the baryon equation
of motions. In particular, all equation of motion terms of the form
\beq
[ \, i v \cdot \nabla , H \,] =-D/F \, (S \cdot u, H)_\pm +
\frac{2}{3}D < S \cdot u \, H > {\bf 1} 
\label{Eqmotion}
\eeq
can be eliminated by appropriate field redefinitions in complete
analogy to the two--flavor case \cite{eckmoj}. The last term in
Eq.(\ref{Eqmotion}) is due to the fact that the baryons are in the
adjoint representation of SU(3). A further reduction in
the number of terms could be achieved by use of the Cayley--Hamilton 
relation, compare Eq.(\ref{CH}). Also, many of the terms given in the
table refer to processes with at least three Goldstone bosons. These
are only relevant in multiple pion or kaon production by photons or
pions off nucleons \cite{bkmpipin}.
The renormalized LECs $d_i^r (\mu)$ are measurable quantities. They
satisfy the renormalization group equations
\beq \label{betai}
\mu \, \frac{d}{d \mu} d_i^r (\mu ) = -\beta_i \,\,\, .
\eeq
Therefore, the choice of another scale $\mu_0$ leads to modified
values of the renormalized LECs,
\beq
 d_i^r (\mu_0 ) =  d_i^r (\mu ) + \beta_i \, \log \frac{\mu}{\mu_0}
 \,\, .
\eeq
We remark that the scale--dependence in the counterterm Lagrangian is,
of course, balanced by the scale--dependence of the renormalized
finite one--loop functional for observable quantities.

\section{A sample calculation}
\label{sec:Ex}

Here, we present a sample calculation to demonstrate the use of 
table~1. Consider the reaction $\gamma (k) \, p (p_1) \to \pi^0 (q) 
\, p (p_2)$ in the threshold region
calculated in SU(3) (for details, see ref.\cite{sven}). The T--Matrix
can be decomposed into S-- and P--wave multipoles as detailed in 
\cite{bkmz}. Whereas in the two--flavor case the loops are all finite
to order $q^3$, in the three--flavor calculation one encounters
divergences already at this order. The transition amplitude has the
form
\begin{eqnarray}
\frac{m}{4\pi W}\,\vec{T}  = 
i\vec{\sigma}\bigg(E_{0+}+\hat{q}\cdot\hat{k}\,P_1\bigg)
+i\vec{\sigma}\cdot\hat{k}\,\hat{q}\,P_2
+\hat{q}\times\hat{k}\,P_3 \,\, ,
\end{eqnarray}
with $W$ the total cms energy, $m$ the nucleon mass
 and the multipoles are complex functions
of the pion cms energy $\omega$. To be specific, consider
the multipole $P_3$. Straightforward evaluation of the irreducible
Feynman graphs  shown in Fig.~3a,b leads to \cite{sven}
\beqa
\label{P3irrs}
P_3^{\rm a} &=&
- C \,\frac{(D+F)^3}{2}\, ( \gamma_3 (\omega) - \gamma_3 (-\omega ) )
= -C \, (D+F)^3\, L \, \omega + {\rm finite}
\nonumber \\
P_3^{\rm b} &=&
C \, D \, \biggl( F^2 -\frac{2}{3}FD - \frac{1}{3}D^2 \biggr)
\, ( \gamma_3^K (\omega) - \gamma_3^K (-\omega ) ) \nonumber \\
&=& C \, D \, \biggl( F^2 -\frac{2}{3}FD - \frac{1}{3}D^2 \biggr) \,
2 L \, \omega + {\rm finite}
\eeqa
with $C = ( e \,|\vec{q} \,|)/(4 \pi F_0^3)$ and 
$\vec{q}$ the three-momentum of the produced neutral pion. The loop functions 
$\gamma_3 (\pm \omega )$ are
given in the review \cite{bkmrev}. Similarly, the reducible graphs
3c,d and their crossed partners lead to
\beqa
\label{P3reds}
P_3^{\rm c} &=&
C \,\frac{(D+F)^3}{2}\, ( \gamma_3 (\omega) - \gamma_3 (-\omega ) )
= C \, (D+F)^3\, L \, \omega + {\rm finite}
\nonumber \\
P_3^{\rm d} &=&
C \, \frac{1}{3} \, (D+F) \, (3F^2+D^2) 
\, ( \gamma_3^K (\omega) - \gamma_3^K (-\omega ) ) \nonumber \\
&=& C \, \frac{1}{3} \, (D+F) \, (3F^2+D^2)  \,
2 L \, \omega + {\rm finite} \,\, .
\eeqa
We now show how these results can be recovered using table~1. The
counterterms $91$, $93$, $94$, $95$ and $96$ lead to 
the structure multiplying $P_3$ for
irreducible graphs. Straightforward calculation leads to
\beqa
\label{P3irrg}
P_3^{\rm irr,ct} &=&
- C \, \biggl( 2 \beta_{91} + \frac{4}{3}
\beta_{93} + 4 \beta_{94} + \frac{4}{3}\beta_{95} + 4 \beta_{96}
\biggr) \, 2 L  \omega  + {\rm finite}   \nonumber \\
&=& C \, \frac{1}{3} (5D^3+13D^2F+3DF^2+3F^3) \, L \, \omega + {\rm finite}
\eeqa
which exactly cancels the divergences from the direct calculation.
The reducible graphs are of course proportional to the equations of
motion and thus the corresponding counterterms are 97 and 98. We
find
\beqa
\label{P3redg}
P_3^{\rm red, ct } &=&
- C \, (D+F) \, \biggl( \frac{1}{3} \beta_{97} + \beta_{98} 
\biggr) \, 2 L \omega  + {\rm finite}    \nonumber \\
&=& -C \, (D+F) \, \biggl( DF + \frac{5D^2+9F^2}{6} \biggr) \, 2 L \, 
\omega + {\rm finite} \,\, ,
\eeqa
in agreement with the direct Feynman graph calculation,
Eq.(\ref{P3reds}). We note that the reducible graphs could be
eliminated using the baryon equations of motion. This would, of
course, also modify the coefficient appearing in Eq.(\ref{P3irrg}).

\section{Summary and conclusions}
\label{sec:summ}

In this paper, we have performed the chiral--invariant renormalization
of the effective three--flavor meson--baryon field theory and
constructed the complete counterterm Lagrangian to leading one--loop 
order $q^3$. To describe the ground state baryon octet, we have used
heavy baryon chiral perturbation theory in the path integral 
formulation \cite{bkkm}. This extends previous work by Ecker \cite{ecker},
who considered the pion--nucleon system, i.e. the two--flavor case.
Since there exist very few explicit calculations in SU(3) where
divergences at ${\cal O}(q^3)$ were evaluated, we list the following
checks on the rather involved manipulations:

\begin{enumerate}

\item[(1)] Reducing our expressions to SU(2) and taking into account
  that $g_A =F+D$ and $F-D=0$, we recover the results of the table in
  ref.\cite{ecker}.
 
\item[(2)] As stressed before, the method destroys covariance in some 
  intermediate steps (see section~6, e.g. Eq.(\ref{coveq}), and the formulae
  in appendix~C). Of course the final results are covariant. This is
  achieved by forming appropriate combinations of the operators
  $O_i^{ab}$, see appendix~C.

\item[(3)] The self--energy contributions $\Sigma_{1,i}^{ab}$ given in
  appendix~C are covariant but not
  hermitian. Hermiticity is restored by again combining 
  appropriate terms. This leads to the complete counterterm
  Lagrangian given in section~7 in terms of the operators
  $\tilde{O}_i^{ab}$, compare table~1.

\item[(4)] Some of the trace relations given in appendix~A are rather
  involved. We have checked these with the help of standard analytical
  packages.

\item[(5)] For the case of pion/kaon photoproduction, we have compared
  our results with the one obtained by direct Feynman graph
  calculation \cite{sven} as detailed in section~8.

\end{enumerate}

In summary, we remark that the method used has the disadvantage of
leading to very lengthy and complicated expressions in the
intermediate steps due to the loss of covariance. There should exist
an improved method which does not share this complication. On the
other hand, this method is very general, i.e. it can be applied to any
differential operator that has the structure given in
Eq.(\ref{S12Vi}}) straightforwardly. In particular, the necessary
inclusion of virtual photons in the pion--nucleon (or meson--baryon)
system can be treated along these lines. We hope to report on the
results of such an investigation in the near future.

\bigskip\bigskip


\section*{Acknowledgments}
We are grateful Sven Steininger for providing us with results
of his calculation prior to publication and for some independent checks.

\bigskip\bigskip\bigskip

\appendix
\def\theequation{\Alph{section}.\arabic{equation}}
\setcounter{equation}{0}
\section{Trace relations and other useful identities}
\label{app:traces}
In this appendix, we collect some useful trace relations used
throughout the text. The completeness relation of the
generators in the Gell-Mann basis reads,
\beq
\lambda^i_{\alpha\beta} \lambda^i_{\alpha' \beta'} = 2\delta_{\alpha\beta'} 
\delta_{\alpha'\beta} - \frac{2}{N} \delta_{\alpha\beta}
\delta_{\alpha'\beta'} \,\,\, .
\eeq
The superscripts '$a,b, \ldots$' refer to the physical basis whereas
the $\lambda_{i,j,\ldots}$ are the generators in the Gell-Mann basis
(called $\lambda_G^i$ in the main text).
Consider now matrices $A,B,C,\ldots \in $ SU(N) (which
are not necessarily traceless). 
\beqa
&&[\lambda_i,[A,\lambda_j]] \, <B\{\lambda_i,\lambda_j \} > =
8 <AB> - 8A <B>  + 8 <A>B \nonumber \\
&& - 4N \{A,B\} \\
&&[\lambda_i,[A,\lambda_j]] \, <B[\lambda_i,\lambda_j ] > = - 4N [A,B]
\\
&&[\lambda_i,[A,\lambda_j]] <[B,\lambda_i][\lambda_j,C]> =
4 \biggl\{ -2B<AC> -2C<AB> - 2A<BC> \nonumber \\
&&
+ <A> \{B,C\} + <B>\{A,C\} 
+ <C> \{A,B\} +  <ABC> + <ACB>  \nonumber \\
&&- N \, ACB - N \, BCA \, \biggr\} \\
&&
A \, \lambda_i \, B \lambda_i \, C = 2AC<B> - \frac{2}{N} ABC \\
&&
X \, \lambda_i \, A [B, \lambda_i] \, Y = 2XY<AB> - 2 XBY <A> \\
&&X \, [A,\lambda_i] \, B [C, \lambda_i] \, Y = \nonumber \\
&& 2XAY<BC> - 2 XACY <B> -2XY<ABC> + 2XCY<AB>
\eeqa
We frequently use 
\beq
< ({\lambda^a}^\dagger , \lambda_i)_\pm (\lambda_i , \lambda^b)_\pm> =
\biggl\{ \biggl( 4N - \frac{8}{N} \biggr)\,D^2 + 4 \, N\,F^2 \biggr\}
<{\lambda^a}^\dagger \, \lambda^b > \,\, .
\eeq
Now we specialize to the case that the matrices $A,B, \ldots$ are
traceless, i.e. $<A>$ = $<B>$ = $\ldots=0$. Useful relations are
\beqa
<({\lambda^a}^\dagger , \lambda_i)_\pm( A, (\lambda^i , \lambda^b)_\pm
)_\pm> &=&
\biggl( 2ND(D^2-F^2) - \frac{8}{N}D^3 \biggr)
<{\lambda^a}^\dagger \, \{ A, \lambda^b \} >  \\
&+& \biggl(- 2NF(D^2-F^2) - \frac{8}{N}D^2F \biggr)
<{\lambda^a}^\dagger \, [ A, \lambda^b ] > \nonumber
\eeqa

\beqa
&&<({\lambda^a}^\dagger , \lambda_i)_\pm( A, (B,(\lambda^i , \lambda^b)_\pm
)_\pm )_\pm > =
2 \bigl( (D +F)^4 + (D-F)^4  \bigr) \,
<{\lambda^a}^\dagger \,  \lambda^b  > <AB> \nonumber \\
&& \qquad\qquad + 4 (D^2-F^2)^2 \, 
\bigl( <{\lambda^a}^\dagger \, A ><B \, \lambda^b >
 + <{\lambda^a}^\dagger \, B ><A \, \lambda^b > \bigr)
\nonumber \\
&&\qquad\qquad + \biggl( N (D^2-F^2)^2 - \frac{8}{N} D^4  \biggr) \,
<{\lambda^a}^\dagger \, \{ A ,\{B, \lambda^b \}\}  > \nonumber \\
&&\qquad\qquad + \biggl( N (D^2-F^2)^2 - \frac{8}{N} D^2 F^2  \biggr) \,
<{\lambda^a}^\dagger \, [ A , [ B, \lambda^b ] ] > \nonumber \\
&&\qquad\qquad -\biggl( \frac{8}{N} D^3 F  \biggr) \, \biggl\{
  <{\lambda^a}^\dagger \, \{ A , [B, \lambda^b ] \}  > 
+ <{\lambda^a}^\dagger \, [ A , \{B, \lambda^b \} ]  > \biggr\}
\eeqa
Other more complicated relations can be derived along similar lines.

To arrive at the terms given in table~1, one makes use of various
relations derived from the Cayley--Hamilton identity,
\beq
A^3 - <A>A^2 + \frac{1}{2} \biggl( <A>^2 - <A^2> \biggr) \, A 
= {\rm det}(A) \,\, \, .
\eeq
For traceless matrices $A_{1,2, \ldots}$ one can derive from that
\beqa
&&\sum_{6 \, {\rm perm.}} <{\lambda^a}^\dagger \, A_1 \, A_2 \, \lambda^b>
-  <{\lambda^a}^\dagger \lambda^b> <A_1 A_2>  \nonumber \\
&& =  <{\lambda^a}^\dagger A_2><A_1 \lambda^b> +  <{\lambda^a}^\dagger
A_1><A_2 \lambda^b>   \,\, ,
\eeqa
which for $A_1=A_2=X$ gives the relation Eq.(\ref{CH}).  Similarly, we
find
\beq
\sum_{24 \, {\rm  perm.}} <{\lambda^a}^\dagger \, A_1 
\, A_2 \, A_3 \, \lambda^b>
 - \sum_{20 \, {\rm perm.}} <{\lambda^a}^\dagger \, \lambda^b><A_1 A_2
 A_3> = 0 \,\,\, .
\eeq
For more relations see \cite{sch}.

\def\theequation{\Alph{section}.\arabic{equation}}
\setcounter{equation}{0}
\section{Products of singular operators}
\label{app:singops}

In this appendix we consider the singularities arising from the
product of the baryon and meson propagators in the evaluation of the
self--energy diagram.
The corresponding singularities can best be extracted
in Euclidean $d$-dimensional Fourier space \cite{jacko}.
There is no difference between SU(2) and SU(3) because the structure 
of divergences is related to the coordinate space and not to the 
flavor space. Before we show how this calculation proceeds, some
general remarks are in order. The $G_n (x,y)$ given in Eq.(\ref{Gxy})
lead to singular terms of the type
\beq
\label{B1}
\frac{1}{ | x - y |^{2\alpha}} \quad .
\eeq
The singular behavior of these terms can be studied by use of the
d--dimensional Fourier-transform,
\beq
\label{Ftrans}
\int d^dx \, \frac{1}{ |x|^{2\alpha}}\, {\rm e}^{i k x } = 
\pi^{d/2} \, \frac{ \Gamma (d/2 -2)}{ \Gamma ( \alpha)} \,
\biggl[ \frac{k}{4} \biggr]^{\alpha -d/2} \,\, ,
\eeq
which has poles at $\alpha-d/2 = 0,1,2, \ldots \,$. In an analogous
manner, we will treat the products of singular operators, related to
the baryon and meson propagators in the self--energy loop. These
have the form $G_n (x,y) \, J_m (x,y)$ and can be treated by 
Fourier--transforms involving now  more complicated integrands of the
type $\exp(-\alpha \,t - \beta /t )$, compare Eqs.(\ref{J0},\ref{Jn}).
To be specific, consider
\beq \label{intGJ}
\int d^dx \, G_n(x) \, J_m (x) \, {\rm e}^{ikx} \,\,\, .
\eeq
To evaluate it, we need a specific representation of the function
$g(x)$ \cite{ecker}
\beq
g(x) = \int \frac{d^d p}{(2\pi)^{d-1}} \, \delta(k \cdot v)  \, 
{\rm e}^{-ipx} \,\, \, .
\eeq
Choosing the Euclidean rest-frame, $v=(0,\ldots,0,1)$, we have $g(x) =
\delta^{d-1} (x)$ and the integral Eq.(\ref{intGJ}) takes the form
\beq \label{int1}
\frac{2}{(4\pi)^{d/2+1/2}}\,\,\Gamma(\alpha)\,\Gamma(\beta)\,
 \int_{-\infty}^\infty dx \, x^{-2\alpha} [ x^2 - 2ikx + \mu^2]^{-\beta}
\eeq
with
\beq
\alpha= \frac{d}{2} -1 - n \,\, , \,\, \beta = 4 -d + 2n + m
-\frac{1}{2} \,\, , \, \,  \alpha+\beta = 2 -  \frac{d}{2} + n + m
+ \frac{1}{2} \,\, . 
\eeq
Only for $m=n=0$ one finds divergences. Using standard methods,
it can be brought into the form
\beq \label{int1a}
\frac{1}{(4\pi)^{2-\epsilon/2}}
\int_{-\infty}^\infty dx \int_0^1 dt \, \frac{(1-t)^{\alpha-1} \,
  t^{\beta  -1} }{ [ x^2 + k^2t^2 + t \mu^2]^{\alpha+\beta} } \,\, .
\eeq
which leads to the divergence
\beq
- \frac{2}{(4\pi)^2} \, \Gamma \biggl( \frac{\ve}{2} \biggr) 
+ {\rm finite} \,\, .
\eeq
Expanding the $\Gamma$ function and going back to coordinate space,
one is finally left with
\beq
- \frac{2}{(4\pi)^2} \,  \frac{2}{\ve} \, \int d^dx \, \delta^d (x) 
+ {\rm finite} \,\, .
\eeq
In a more transparent way one can evaluate the integral (\ref{intGJ})
by transforming it into typical heavy baryon one--loop integrals in
momentum space.
Straightforward algebra leads to   
the following list \cite{ecker} that contains
all singular products appearing in $\Sigma_1$ (with $\ve = 4-d$):
\beqa
G_0(x-y) J_0(x-y) &\sim& - \frac{2}{(4\pi)^2v^2} \frac{2}{\ve} \delta^d(x-y)\\
G_0(x-y)v \cdot \partial J_0(x-y) &\sim&  \frac{2}{(4\pi)^2v^2}
\frac{2}{\ve} 
v \cdot \partial \delta^d(x-y) \\
S_\mu S_\nu \partial_\mu \partial_\nu G_0(x-y) J_0(x-y) &\sim&
 \frac{2}{(4\pi)^2}  \frac{2}{\ve}
S_\mu S_\nu \delta_{\mu\nu} (v \cdot \partial)^2
\delta^d(x-y) \\
S_\mu S_\nu \partial_\mu \partial_\nu G_0(x-y)v \cdot \partial J_0(x-y) &\sim&
- \frac{2}{3(4\pi)^2}  \frac{2}{\ve}
S_\mu S_\nu \delta_{\mu\nu} (v \cdot \partial)^3
\delta^d(x-y) \\
S_\mu S_\nu \partial_\mu \partial_\nu G_0(x-y) J_1(x-y) &\sim&
- \frac{2}{3(4\pi)^2} \frac{2}{\ve}
 S_\mu S_\nu \delta_{\mu\nu} \delta^d(x-y) \\
S_\mu S_\nu \partial_\mu \partial_\nu G_0(x-y)v \cdot \partial  J_1(x-y) &\sim&
 \frac{2}{(4\pi)^2}  \frac{2}{\ve}
S_\mu S_\nu \delta_{\mu\nu} v \cdot \partial
\delta^d(x-y) \\
S_\mu S_\nu \partial_\mu \partial_\nu G_1(x-y) J_0(x-y) &\sim&
 \frac{1}{(4\pi)^2v^2}  \frac{2}{\ve}
S_\mu S_\nu \delta_{\mu\nu} \delta^d(x-y) \\
S_\mu S_\nu \partial_\mu \partial_\nu G_1(x-y)v \cdot \partial J_0(x-y) &\sim&
- \frac{1}{(4\pi)^2v^2}  \frac{2}{\ve}
S_\mu S_\nu \delta_{\mu\nu} v \cdot \partial
\delta^d(x-y) .
\eeqa
We explicitly make use of the identity $S \cdot v = 0$. In the case
of a more general differential operator acting on the meson propagator
we get a richer structure of divergences.

\def\theequation{\Alph{section}.\arabic{equation}}
\setcounter{equation}{0}
\section{Contributions to the self--energy}
\label{app:selfen}

Here we list the  operators corresponding to the three cases discussed
in section~6 and the resulting contribution to the divergent part
of the self--energy functional. 

\beqa
O_1^{ab} &=& V_i^{ac(1)} \biggl\{ v \cdot \gamma^{ij} \, V_j^{cb(1)}
+ v \cdot d^{cd} \, V_i^{db(1)} + V_i^{cb(1)} \,v \cdot \partial
\biggr \} \nonumber \\
\Sigma_{1,1}^{ab} &=& \frac{i}{4} \, \biggl\{ 2 <{\lambda^a}^\dagger
\, [v \cdot \nabla ,\lambda^b]> <(v \cdot u)^2>+ 4 <
{\lambda^a}^\dagger \, v \cdot u>< v \cdot u \, [v \cdot \nabla ,
\lambda^b]>  \nonumber \\
&& + 2 <{\lambda^a}^\dagger \lambda^b > < v \cdot u \, [v \cdot \nabla
, v \cdot u]> +2 <{\lambda^a}^\dagger \, v \cdot u>
< [v \cdot \nabla , v \cdot u] \, \lambda^b> \nonumber \\
&&+2<{\lambda^a}^\dagger \, [v \cdot \nabla , v \cdot u]><v \cdot u \, 
\lambda^b> + 3 <{\lambda^a}^\dagger \, \{ (v\cdot u)^2 , [v \cdot
\nabla, \lambda^b ] \}> \nonumber \\
&& + 3 <{\lambda^a}^\dagger \,  v\cdot u \, [v \cdot\nabla, v \cdot
u] \, \lambda^b > + 3 <{\lambda^a}^\dagger \, \lambda^b \,
[v \cdot\nabla, v \cdot u] \, v \cdot u > \biggr\}
\eeqa

\beqa
O_2^{ab} &=& V_i^{ac(1)} < {\lambda^c}^\dagger \, (S \cdot u , 
\lambda^d)_\pm \,>  \, V_i^{db(1)}
\nonumber \\
\Sigma_{1,2}^{ab} &=& -\frac{1}{2}D/F \, \biggl[ \,
<{\lambda^a}^\dagger \, (v \cdot u, \lambda^b)_\pm \,><v \cdot u \, S \cdot
u> \nonumber \\ && \qquad\qquad -
<{\lambda^a}^\dagger \, (S\cdot u ,v \cdot u)_\pm> <v \cdot u \,
 \lambda^b\ > \nonumber \\
&& - <{\lambda^a}^\dagger \, v \cdot u> < (v \cdot u
, S \cdot u )_\pm \, \lambda^b> -  \frac{1}{2} \, 
<{\lambda^a}^\dagger \, (S\cdot u , \lambda^b )_\pm \,> <(v \cdot
u)^2> \biggr] \nonumber \\ && + \frac{3}{4}(D+F)\,  <{\lambda^a}^\dagger \, 
S \cdot u \, \lambda^b \, (v \cdot u)^2 \,> + \frac{3}{4}(D-F)\,
<{\lambda^a}^\dagger \, (v \cdot u)^2 \, \lambda^b \, S \cdot u \, >
\nonumber \\ && +\frac{D}{2}\, \biggl[ \, <{\lambda^a}^\dagger \, \lambda^b \,>
<(v \cdot u)^2 \, S\cdot u> - <{\lambda^a}^\dagger \, (v \cdot u)^2>
< S\cdot u \,  \lambda^b \,> \nonumber \\
&& \qquad \qquad \qquad \qquad \qquad \qquad \qquad
- <{\lambda^a}^\dagger \, S\cdot u >
<(v \cdot u)^2 \,  \lambda^b \,> \, \biggr]
\eeqa

\beqa
O_3^{ab} &=& V_i^{ac(1)} \, {(\gamma^{\mu \nu})}^{ij} \,
 \, V_j^{cb(2)} + ( 1 \leftrightarrow 2)
\nonumber \\
\Sigma_{1,3}^{ab} &=& -\frac{3}{2}D/F \, <{\lambda^a}^\dagger \,
( [ \Gamma^{\mu \nu} , v \cdot u ] , \lambda^b \,)_\pm \,> 
\, S_\mu \, v_\nu
\eeqa

\beqa
O_4^{ab} &=& V_i^{ac(2)} \, \biggl\{ \delta^{ij} \, {(v \cdot
  d)^3\,}^{cd} + 3 v \cdot \gamma^{ij} {(v \cdot d)^2\,}^{cd} + 3 v \cdot d
\, v \cdot \gamma^{ij} \, (v \cdot d)^{cd} \nonumber \\
&& \qquad\qquad\qquad\qquad\qquad + (v \cdot d^2 v \cdot \gamma)^{ij}
\biggr\} \,   \, V_j^{cb(2)} \nonumber \\
\Sigma_{1,4}^{ab} &=& -i\biggl(\frac{10}{3}D^2 + 6 F^2 \biggr) \,
 <{\lambda^a}^\dagger \, [v \cdot \nabla, [v \cdot \nabla, [v \cdot
 \nabla , \lambda^b]]] \, >
\eeqa

\beqa
O_{5}^{ab} &=& V_i^{ac(2)} \,\biggl[ (\gamma_{\mu\nu} \, v \cdot 
\gamma)^{ij} \, \delta^{cd} + \gamma_{\mu\nu}^{ij} \, v \cdot d^{cd}
 \, \biggr] \,V_j^{db(2)}  \nonumber \\
\Sigma_{1,5}^{ab} &=& S^\mu \,\biggl[
\biggl( \frac{10}{3}D^2 + 6 F^2 \biggr) \, 
<{\lambda^a}^\dagger \, [\Gamma_{\mu\nu} , [v \cdot \nabla, \lambda^b]]\,>
\nonumber \\ && \qquad \qquad \quad
+ 12DF \, <{\lambda^a}^\dagger \, \{ \Gamma_{\mu\nu} , [v \cdot \nabla, 
\lambda^b] \} \,> \biggr] \, S^\nu  
\eeqa

\beqa
O_{6}^{ab} &=& V_i^{ac(2)} \,\biggl[ \delta_{\mu \nu} ( d_\lambda 
\gamma_{\lambda\kappa})^{ij} \, v^\kappa + 2(v \cdot \gamma_{\mu\nu})^{ij}
+ 2 (d_\nu \gamma_{\mu\kappa})^{ij} \, v^{\kappa} \, \biggr] 
\,V_j^{cb(2)}  \nonumber \\ 
\Sigma_{1,6}^{ab} &=&i \,\biggl[
\biggl( -\frac{25}{36}D^2 - \frac{5}{4} F^2 \biggr) \, 
<{\lambda^a}^\dagger \, [[\nabla^\mu , \Gamma_{\mu\nu} v^\nu], \lambda^b]\,>
\nonumber \\ && \qquad \qquad \quad
-\frac{5}{2}DF \, <{\lambda^a}^\dagger \, \{  \nabla^\mu , 
\Gamma_{\mu\nu} v^\nu ], \lambda^b \} \,> \nonumber \\ 
&& + i \, \epsilon^{\mu\nu\rho\sigma} \, v_\rho \, S_\sigma \, \biggl\{
3DF \, <{\lambda^a}^\dagger \, \{ [ v \cdot  \nabla, 
\Gamma_{\mu\nu}  ], \lambda^b \} \,> \nonumber \\ 
&&  \qquad \quad + \biggl( \frac{5}{6}D^2 + \frac{3}{2} F^2 \biggr) \,
<{\lambda^a}^\dagger \, [ [ v \cdot  \nabla, \Gamma_{\mu\nu} ], 
\lambda^b ] \,> \, \biggr\} \biggr]  
\eeqa

\beqa
O_{7}^{ab} &=& V_i^{ac(2)} \,\biggl[ 2 v \cdot \gamma^{ij} \, 
( v \cdot d)^{cd}  
+ <{\lambda^c}^\dagger \, ( S\cdot u , \, \lambda^e )_\pm > 
\, {(v \cdot d)^2}^{ed} \, \delta^{ij} \nonumber \\
&& \qquad \qquad \qquad + (v \cdot d \, v \cdot \gamma)^{ij}
<{\lambda^c}^\dagger \, ( S\cdot u , \, \lambda^d )_\pm >
 \biggr] \,V_j^{db(2)} \nonumber \\
\Sigma_{1,7}^{ab} &=& -(D^3 + 3 D F^2) \,
<{\lambda^a}^\dagger \, \{ S \cdot u , [ v \cdot \nabla , [ v \cdot \nabla,
\, \lambda^b\} ] \} > \nonumber \\
&& + \biggl( 3 F^3 - \frac{5}{3} D^2 F \biggr) \,
<{\lambda^a}^\dagger \,[ S \cdot u , [ v \cdot \nabla, [ v \cdot
\nabla , \, \lambda^b] ] ] > 
\eeqa

\beqa
O_{8}^{ab} &=& V_i^{ac(2)} \,\biggl[  v \cdot \gamma^{ij} \, a_1^{cd}  
+ (a_1 \, v \cdot d)^{cd} \delta^{ij} \biggr] \,V_j^{db(2)} \nonumber \\
\Sigma_{1,8}^{ab} &=& - D( 3 F^2 + D^2) \,
<{\lambda^a}^\dagger \, \{ [ v \cdot \nabla, S \cdot u] ,  [ v \cdot
\nabla , \, \lambda^b] \} > \nonumber \\ 
&& + \biggl( 3 F^3 - \frac{5}{3} D^2 F \biggr) \,
<{\lambda^a}^\dagger \, [ [ v \cdot \nabla, S \cdot u] ,  [ v \cdot
\nabla , \, \lambda^b] ] > 
\eeqa

\beqa
O_{9}^{ab} &=& V_i^{ac(2)} \, [  v \cdot d \, , \, \, a_1 ]^{cd}  
\, \,V_i^{db(2)} \nonumber \\
\Sigma_{1,9}^{ab} &=& -\frac{1}{3} D (D^2  + 3 F^2 ) \,
<{\lambda^a}^\dagger \, \{ [ v \cdot \nabla,[ v \cdot \nabla, S \cdot u]] ,  
\, \lambda^b \} > \nonumber \\ 
&& - \frac{1}{9} F ( 5D^2 - 9F^2 ) \,
<{\lambda^a}^\dagger \, [ [ v \cdot \nabla, [ v \cdot \nabla,S \cdot u]] , 
\, \lambda^b ]  > 
\eeqa

\beqa
O_{10}^{ab} &=& V_i^{ac(2)} \, \biggl[ \gamma_{\mu \nu}^{ij} \,
<{\lambda^a}^\dagger \, (S \cdot u, \lambda^d)_\pm \, >
\biggr] \, V_j^{db(2)} 
\nonumber \\
\Sigma_{1,10}^{ab} &=& S^\mu \, \biggl\{
-4F(F^2+3D^2) \,<{\lambda^a}^\dagger \, \lambda^b\,>
< S \cdot u \, \Gamma_{\mu\nu}> \nonumber \\
&& +4\,F\,(D^2-F^2) \,<{\lambda^a}^\dagger \,
\Gamma_{\mu\nu}> < S \cdot u \,  \lambda^b\,> \nonumber \\
&& + 4\,F\,(D^2-F^2) \, \,<{\lambda^a}^\dagger \, S \cdot u>
<\Gamma_{\mu\nu} \, \lambda^b\,> \nonumber \\
&& -6\,(D-F)\,(D^2-F^2) \, <{\lambda^a}^\dagger \, S \cdot u \,
\lambda^b\,  \Gamma_{\mu\nu}\,> \nonumber \\
&& +6\,(D+F)\,(D^2-F^2) \, <{\lambda^a}^\dagger \,  \Gamma_{\mu\nu}\, 
\lambda^b\, S \cdot u \, >  \\
&& -\frac{8}{3}D^2 \, D/F\, \biggl[ <{\lambda^a}^\dagger \,
[\Gamma_{\mu\nu}, (\lambda^b,S\cdot u)_\pm]> +  <{\lambda^a}^\dagger
\,( [ \Gamma_{\mu\nu}, \lambda^b], S\cdot u)_\pm>\biggr]\biggr\} \, S^\nu
\nonumber
\eeqa 

\beqa
O_{11}^{ab} &=& V_i^{ac(2)} \,\biggl[ (\sigma \, v \cdot 
\gamma)^{ij} \, \delta^{cd} + \sigma^{ij} \, v \cdot d^{cd}
 \, \biggr] \,V_j^{db(2)} 
\nonumber \\
\Sigma_{1,11}^{ab} &=& i \, \biggl\{
\biggl( \frac{3}{8}D^2-\frac{9}{8}F^2 \biggr) \, 
<{\lambda^a}^\dagger \, \{ \chi_+ , [v \cdot \nabla, \lambda^b]\}\,>
-\frac{5}{4}DF \, <{\lambda^a}^\dagger \, [ \chi_+ , [v \cdot \nabla, 
\lambda^b]]\,> \nonumber \\
&&  -\biggl(\frac{13}{12}D^2+\frac{3}{4}F^2 \biggr) \,
<{\lambda^a}^\dagger \, [v \cdot \nabla, \lambda^b]\,><\chi_+>
+\frac{1}{2}D^2 \, <{\lambda^a}^\dagger \, [u_\mu,[u^\mu,[v \cdot 
\nabla, \lambda^b]]]> \nonumber \\
&&   -\frac{3}{4}(D^2+F^2)  \,
<{\lambda^a}^\dagger \, [v \cdot \nabla, \lambda^b]\,><u^2>
- \frac{9}{8}(D^2 + F^2 ) 
\, <{\lambda^a}^\dagger \, \{ u^2 , \,[v \cdot 
\nabla, \lambda^b]\}> \nonumber \\
&&- \frac{9}{4}D F  \, <{\lambda^a}^\dagger \, [ u^2 , \,[v \cdot 
\nabla, \lambda^b]]> + \frac{3}{2}(D^2 - F^2) \, <{\lambda^a}^\dagger \,
u_\mu>< u^\mu \, [v \cdot \nabla, \lambda^b] \,> \biggr\} 
\nonumber \\ && \,
\eeqa

\beqa
O_{12}^{ab} &=& V_i^{ac(2)} \, [ v \cdot d, \, \sigma ]^{ij} \, V_j^{cb(2)} 
\nonumber \\
\Sigma_{1,12}^{ab} &=& i \biggl\{ \frac{3}{16}(D^2 -3F^2) \, 
<{\lambda^a}^\dagger \, \{ [v \cdot \nabla,\chi_+] ,  \lambda^b]\} >
-\frac{5}{8}DF \, <{\lambda^a}^\dagger \, [ [v \cdot \nabla,\chi_+] ,  
\lambda^b] > \nonumber \\
&&  -\frac{1}{24}(13D^2 + 9 F^2 ) \,
<{\lambda^a}^\dagger \,  \lambda^b\,><[v \cdot \nabla,\chi_+]>
\nonumber \\
&& -\frac{3}{4}(D^2 + F^2)\,<{\lambda^a}^\dagger \,  \lambda^b\,> <u_\mu \,
[v \cdot \nabla,u^\mu]> \nonumber \\
&&+\frac{1}{4}D^2 \,\biggl( <{\lambda^a}^\dagger \, [u_\mu,[[v \cdot 
\nabla, u^\mu], \lambda^b]]> +  <{\lambda^a}^\dagger \,[[ v \cdot
\nabla, u_\mu], [u^\mu, \lambda^b]]> \biggr) \nonumber \\
&&-\frac{9}{16}(D^2 + F^2) \,
 <{\lambda^a}^\dagger \, \{ \{ [v \cdot \nabla , u_\mu] \,  
, u^\mu\} ,\, \lambda^b \} > \nonumber \\
&&-\frac{9}{8}DF \,  <{\lambda^a}^\dagger \,[ \{  [ v \cdot
\nabla, u_\mu], u^\mu\} \, \lambda^b ] \,> \\
&& + \frac{3}{4}(D^2 - F^2) \biggl( \,
<{\lambda^a}^\dagger \, [v \cdot \nabla, u_\mu]><u^\mu \, \lambda^b >
+ <{\lambda^a}^\dagger \,  u_\mu>< [v \cdot \nabla, u^\mu]\,
\lambda^b > \biggr) \biggr\} \nonumber
\eeqa

\beqa
O_{13}^{ab} &=& V_i^{ac(2)} \, \biggl[ \sigma^{ij} \,
<{\lambda^c}^\dagger \, (S \cdot u, \lambda^d)_\pm \, >
\biggr] \, V_j^{db(2)} 
\nonumber \\
\Sigma_{1,13}^{ab} &=&
\frac{2}{9}D^3 + \frac{1}{4}D(D^2-F^2) 
\, \biggl( <{\lambda^a}^\dagger \, \chi_+>
< S \cdot u \, \lambda^b\,> + <{\lambda^a}^\dagger \, S \cdot u>
<\chi_+ \, \lambda^b\,>\biggr)  \nonumber \\
&+&  \biggl( \frac{1}{9}D^2 \, D/F + \frac{1}{8}(D^2-F^2) \, D/(-F) \biggr)\,
<{\lambda^a}^\dagger \,( S \cdot u \, , \lambda^b)_\pm\,> <\chi_+>
\nonumber \\
&+&  \frac{1}{4}D (D^2 + 3F^2) \, \,<{\lambda^a}^\dagger \,
\lambda^b\,>  <S \cdot u \, \chi_+ \,>
\nonumber \\
&-& \biggl(\frac{1}{3}D^3 -\frac{3}{32}D(D^2-F^2)\biggr) 
\biggl[ <{\lambda^a}^\dagger \, \{ \chi_+, \{ S\cdot u ,
\lambda^b \}\}> +  <{\lambda^a}^\dagger \, \{ S\cdot u , \{ \chi_+ ,
\lambda^b \}\}> \biggr]  \nonumber \\
&-& \biggl( \frac{3}{32} F \, (D^2-F^2) \biggr) 
\biggl[ <{\lambda^a}^\dagger \, \{ \chi_+, [ S\cdot u ,
\lambda^b ] \}> +  <{\lambda^a}^\dagger \, [ S\cdot u , \{ \chi_+ ,
\lambda^b \} ] > \biggr]  \nonumber \\
&-& \biggl(\frac{1}{6}D F^2 +\frac{3}{32} D (D^2-F^2) \biggr) 
\biggl[ <{\lambda^a}^\dagger \, [ \chi_+, [ S\cdot u ,
\lambda^b ] ] > +  <{\lambda^a}^\dagger \, [ S\cdot u , [ \chi_+ ,
\lambda^b ] ] > \biggr]  \nonumber \\
&-& \biggl(\frac{1}{6}D^2 F -\frac{3}{32} F (D^2-F^2) \biggr) 
\biggl[ <{\lambda^a}^\dagger \, [ \chi_+, \{ S\cdot u ,
\lambda^b \} ] > +  <{\lambda^a}^\dagger \, \{ S\cdot u , [ \chi_+ ,
\lambda^b ] \} > \biggr]  \nonumber \\
&+& \frac{1}{8} D/(-F) (D^2-F^2)   <{\lambda^a}^\dagger (  S \cdot u,
\lambda^b )_\pm > <u^2>  \nonumber \\
&+& \frac{D}{4} (D^2 + 3 F^2)  <{\lambda^a}^\dagger \,  \lambda^b  >
< S \cdot u \, u \cdot u>   \\
&-&  \frac{D}{4} (D^2 + 3F^2)  <{\lambda^a}^\dagger \{ u_\mu ,  \lambda^b \}  >
< u^\mu \, S \cdot u >  \nonumber \\
&-&  \frac{F}{4} (3D^2 + F^2)  <{\lambda^a}^\dagger [ u_\mu ,  \lambda^b ]  >
< u^\mu \, S \cdot u > \nonumber \\
&-& \frac{1}{4} D/F (D^2 - F^2) \biggl[
<{\lambda^a}^\dagger \, u_\mu>< ( u^\mu , \, S
\cdot u )_\pm   \lambda^b >  + <{\lambda^a}^\dagger 
( S\cdot u , \, u_\mu )_\pm ><  u^\mu   \lambda^b > \biggr] \nonumber \\
&+& \frac{D}{4} (D^2-F^2) 
\biggl[ <{\lambda^a}^\dagger \,  S \cdot u >< u \cdot u \,
\lambda^b  > + <{\lambda^a}^\dagger \,  u \cdot u >< S \cdot u \,
\lambda^b  > \biggr] \nonumber \\
&-& \frac{D^2}{6} D/(-F) \biggl[ <{\lambda^a}^\dagger [ u_\mu , [  u^\mu , ( S
\cdot u , \lambda^b )_\pm  ] ] > +  
 <{\lambda^a}^\dagger ( S \cdot u , [ u_\mu , [  u^\mu , \lambda^b  ]
 ])_\pm  >  \biggr] \nonumber \\
&+&  \frac{3}{8} (D-F)(D^2 - F^2) 
 <{\lambda^a}^\dagger \,S \cdot u \,  \lambda^b \, u \cdot u > \nonumber \\
&+&  \frac{3}{8} (D+F)(D^2 - F^2) 
 <{\lambda^a}^\dagger \, u \cdot u \,  \lambda^b \, S \cdot u > \nonumber 
\eeqa

\beqa
O_{14}^{ab} &=& V_i^{ac(2)} \,\biggl[ (a_2 \, v \cdot d)^{cd} \, \delta^{ij} 
+  v \cdot \gamma^{ij} \,  a_2^{cd} \, \biggr] \,V_j^{db(2)} 
\nonumber \\
\Sigma_{1,14}^{ab} &=& -2 S^\mu S^\kappa S^\tau S_\mu \biggl\{
-\frac{16}{3}D^2  (D/F)^2\,
<{\lambda^a}^\dagger \, ( [i \, v \cdot \nabla , \lambda^b] \, , \, (u_\kappa,
u_\tau)_\pm)_\pm  >  \nonumber \\
&& +\biggl( \frac{32}{9}D^4 + 4(D^4 + 6D^2F^2 + F^4) \biggr) \, 
<{\lambda^a}^\dagger \,  [i \, v \cdot \nabla , \lambda^b]>< u_\kappa
u_\tau >
\nonumber \\ 
&& +\biggl( \frac{32}{9}D^4 +4 (D^2-F^2)^2 \biggr) \, 
<{\lambda^a}^\dagger \, u_\kappa \,><   u_\tau \,
[i \, v \cdot \nabla , \lambda^b]> \nonumber \\ 
&& +\biggl( 4 (D^2-F^2)^2 \biggr) \, 
<{\lambda^a}^\dagger \, u_\tau \,><   u_\kappa \,
[i \, v \cdot \nabla , \lambda^b]> \nonumber \\ 
&& +\biggl( 6(D^2-F^2)^2 - \frac{16}{3}D^2  (D^2+F^2) \biggr) \, 
<{\lambda^a}^\dagger \, u_\kappa  \,  u_\tau  \,
[i \, v \cdot \nabla , \lambda^b]\,> \nonumber \\ 
&& +\biggl( 6(D^2-F^2)^2 - \frac{16}{3}D^2  (D^2+F^2) \biggr) \, 
<{\lambda^a}^\dagger \,
[i \, v \cdot \nabla , \lambda^b] \,  u_\tau  \,  u_\kappa  \,> \nonumber \\ 
&& +\biggl(  - \frac{16}{3}D^2  (D^2-F^2) \biggr) \, 
<{\lambda^a}^\dagger \, u_\kappa \,
[i \, v \cdot \nabla , \lambda^b]   \,  u_\tau  \,> \nonumber \\ 
&& +\biggl(  - \frac{16}{3}D^2  (D^2-F^2) \biggr) \, 
<{\lambda^a}^\dagger \, u_\tau \,
[i \, v \cdot \nabla , \lambda^b]   \,  u_\kappa  \,> 
\biggr\}
\eeqa

\beqa
O_{15}^{ab} &=& V_i^{ac(2)} \,\biggl[  
<{\lambda^c}^\dagger \, ( S\cdot u , \, \lambda^e )_\pm > \, a_1^{ed} 
+ [v \cdot d, a_2 ]^{cd} \, \biggr] \,V_i^{db(2)} 
\nonumber \\
\Sigma_{1,15}^{ab} &=&-\frac{2}{3} i \, 
 S^\mu S^\kappa S^\tau S_\mu \, \biggl\{
-\frac{32}{3}D^2 \, (D/F)^2
 \,< ({\lambda^a}^\dagger \, , \lambda^b)_\pm \, ( u_\kappa ,  
[v \cdot \nabla, u_\tau] )_\pm > \nonumber \\
&& -\frac{16}{3}D^2 \, (D/F)^2
 \,< ({\lambda^a}^\dagger \, , \lambda^b)_\pm \,  
( [ v \cdot \nabla,  u_\kappa] , u_\tau )_\pm > \nonumber \\
&&+ \biggl( \frac{64}{9}D^4 + 8(D^4 + 6D^2F^2 +F^4) \biggr)
 \,< {\lambda^a}^\dagger \,  \lambda^b> < u_\kappa \,  
[v \cdot \nabla, u_\tau]  > \nonumber \\
&&+ \biggl( \frac{32}{9}D^4 + 4(D^4 + 6D^2F^2 +F^4) \biggr)
 \,< {\lambda^a}^\dagger \,  \lambda^b>   
<[v \cdot \nabla, u_\kappa] \, u_\tau > \nonumber \\
&&+ \biggl( \frac{64}{9}D^4 + 8(D^2-F^2)^2 \biggr)
 \,< {\lambda^a}^\dagger \, u_\kappa><   
[v \cdot \nabla, u_\tau] \,  \lambda^b > \nonumber \\
&& + 8(D^2-F^2)^2 
 \,< {\lambda^a}^\dagger \,  [v \cdot \nabla, u_\tau] >
<u_\kappa \,  \lambda^b > \nonumber \\
&&+ \biggl( \frac{32}{9}D^4 + 4(D^2-F^2)^2 \biggr)
 \,< {\lambda^a}^\dagger \,  [v \cdot \nabla, u_\kappa] >
<u_\tau \,  \lambda^b > \nonumber \\
&&+  4 (D^2-F^2)^2 
 \,< {\lambda^a}^\dagger \, u_\tau>   
<[v \cdot \nabla, u_\kappa] \,  \lambda^b > \nonumber \\
&&+ \biggl( -\frac{32}{3}D^2 (D^2+F^2) + 12(D^2-F^2)^2 \biggr)
 \,< {\lambda^a}^\dagger \, u_\kappa \,   
[v \cdot \nabla, u_\tau] \,  \lambda^b > \nonumber \\
&&+ \biggl(- \frac{16}{3}D^2(D^2+F^2) + 6 (D^2-F^2)^2 \biggr)
 \,< {\lambda^a}^\dagger \,  [v \cdot \nabla, u_\kappa] \,
u_\tau \,  \lambda^b > \nonumber \\
&&+ \biggl( -\frac{32}{3}D^2 (D^2+F^2) + 12(D^2-F^2)^2 \biggr)
 \,< {\lambda^a}^\dagger \, \lambda^b\,    
[v \cdot \nabla, u_\tau] \, u_\kappa > \nonumber \\
&&+ \biggl( -\frac{16}{3}D^2 (D^2+F^2) + 6 (D^2-F^2)^2 \biggr)
 \,< {\lambda^a}^\dagger \, \lambda^b\,    
u_\tau \,[v \cdot \nabla,  u_\kappa ] > \nonumber \\
&& -\frac{32}{3}D^2 (D^2-F^2)  \,< {\lambda^a}^\dagger \,  u_\kappa 
\lambda^b\, [v \cdot \nabla,  u_\tau  ] > \nonumber \\
&& -\frac{16}{3}D^2 (D^2-F^2)  \,< {\lambda^a}^\dagger \,  
 [v \cdot \nabla,  u_\kappa ] \lambda^b\, u_\tau > \nonumber \\
&& -\frac{32}{3}D^2 (D^2-F^2)  \,< {\lambda^a}^\dagger \,  
 [v \cdot \nabla,  u_\tau ] \lambda^b\, u_\kappa > \nonumber \\
&& -\frac{16}{3}D^2 (D^2-F^2)  \,< {\lambda^a}^\dagger \,  u_\tau 
\lambda^b\, [v \cdot \nabla,  u_\kappa  ] > 
\eeqa

\beqa
&& O_{16}^{ab} = V_i^{ac(2)} \,\biggl[  
<{\lambda^a}^\dagger \, ( S\cdot u , \, \lambda^d )_\pm > 
\, a_2^{de} \, \biggr] \,V_i^{eb(2)} 
\nonumber \\
&& \Sigma_{1,16}^{ab} = \nonumber \\ 
&& - \frac{2}{3} \, S^\kappa \,  \biggl\{
\biggl[  S^\mu \,  \, S^\nu \, S^\tau \biggr]
 \biggl( D^5 + 10D^3 F^2 + 5 D F^4 \biggr) 
\,<{\lambda^a}^\dagger \, \lambda^b > 
< \{ u_\mu , \, u_\nu \} \, u_\tau >  \nonumber \\
&&+ \biggl[ S^\mu   \, S^\nu  \, S^\tau \biggr]
 \biggl( 5D^4F + 10D^2 F^3 +   F^5 \biggr) 
\,<{\lambda^a}^\dagger \, \lambda^b > 
< [ u_\mu , \, u_\nu ] \, u_\tau >  \nonumber \\
&&
+\biggl[  S^\mu   \, S^\nu \, S^\tau \biggr]
\biggl( 6(D+F) (D^2 - F^2)^2 -\frac{8}{3}D^2 (D+F)^3 \biggr)
\,<{\lambda^a}^\dagger \, u_\mu \, u_\nu \,  u_\tau  \, \lambda^b > 
\nonumber \\
&&+\biggl[  S^\tau   \, S^\nu \, S^\mu \biggr]
\biggl( 6(D-F) (D^2 - F^2)^2 -\frac{8}{3}D^2 (D-F)^3 \biggr)
\,<{\lambda^a}^\dagger \, \lambda^b \, u_\mu \, u_\nu \,  u_\tau   > 
\nonumber \\
&&+\biggl[  S^\mu \,\{ S^\nu , S^\tau \}  + S^\tau
\, S^\mu \, S^\nu \biggr]
\biggl( -\frac{8}{3}D^2 (D+F) (D^2-F^2) \biggr)
\,<{\lambda^a}^\dagger \, u_\mu \, u_\nu \, \lambda^b \,  u_\tau   > 
\nonumber \\
&&+\biggl[  S^\mu \, S^\tau \, S^\nu + S^\tau \, \{ S^\mu , S^\nu \}  \biggr]
\biggl( -\frac{8}{3}D^2 (D-F) (D^2-F^2) \biggr)
\,<{\lambda^a}^\dagger \, u_\mu \, \lambda^b \, u_\nu \,  u_\tau  \, > 
\nonumber \\
&& +\biggl[  S^\nu \, \{ S^\tau , S^\mu  \}  + S^\mu
\, S^\nu \, S^\tau \biggr]
\,\biggl( 2 (D+F)^4 (D-F)\biggr) 
\,<{\lambda^a}^\dagger \lambda^b\, u_\mu   > 
< u_\nu \, u_\tau > \nonumber \\
&& +\biggl[ \{ S^\nu , S^\mu \} \, S^\tau + S^\tau \, S^\nu \, S^\mu \biggr]
\,\biggl( 2 (D+F) (D-F)^4 \biggr) 
\,<{\lambda^a}^\dagger \, u_\mu \, \lambda^b  > 
< u_\nu \, u_\tau > \nonumber \\
&& +\biggl[  S^\mu \,\{ S^\nu , S^\tau \}  + S^\tau
\, S^\mu \, S^\nu \biggr]
\,\biggl( 2 (D+F) (D^2-F^2)^2 \biggr) 
\,<{\lambda^a}^\dagger \, u_\mu \, u_\nu > 
< u_\tau \lambda^b > \nonumber \\
&& +\biggl[  S^\tau \, S^\nu \, S^\mu + S^\nu \, \{ S^\tau , S^\mu \} \biggr]
\,\biggl( 2 (D-F) (D^2-F^2)^2 \biggr) 
\,<{\lambda^a}^\dagger \, u_\mu \, u_\nu > 
< u_\tau \lambda^b > \nonumber \\
&& +\biggl[  S^\nu \, \{ S^\tau , S^\mu \}  + S^\mu
\, S^\nu \, S^\tau \biggr]
\,\biggl( 2 (D+F) (D^2-F^2)^2 \biggr) 
\,<{\lambda^a}^\dagger \, u_\mu  > 
< u_\nu \, u_\tau \lambda^b > \nonumber \\
&& +\biggl[  S^\mu \, S^\tau \, S^\nu +  S^\tau \, \{ S^\mu , S^\nu \} \biggr]
\,\biggl( 2 (D-F) (D^2-F^2)^2 \biggr) 
\,<{\lambda^a}^\dagger \, u_\mu  > 
< u_\nu \, u_\tau \lambda^b > \biggr\} \, S_\kappa \nonumber \\
&& - \frac{2}{3} \, 
S^\kappa\,  S^\mu \, S^\nu \,  S^\tau \,  S_\kappa \, \biggl\{ 
-\frac{8}{3}D^2 (D/F)^3
< ( {\lambda^a}^\dagger \, , \lambda^b )_\pm \, ( u_\mu ,\, (u_\nu ,
u_\tau)_\pm )_\pm  >  \nonumber \\
&&  
-\frac{8}{3}D^2 (D/F)^3
< ( {\lambda^a}^\dagger \, , \lambda^b )_\pm \, ( (u_\mu , u_\nu )_\pm
, \, u_\tau)_\pm  >  \nonumber \\
&&  
+\frac{32}{9}D^4 (D/F)
< {\lambda^a}^\dagger \,  \lambda^b > <  u_\mu  \, ( u_\nu ,\,
u_\tau)_\pm  >  \nonumber \\
&&
-\frac{8}{3}D^2 (D/F)^3
< ( u_\mu , \, {\lambda^a}^\dagger)_\pm  \, ( \lambda^b ,\,  (u_\nu ,
u_\tau )_\pm )_\pm  >  \nonumber \\ 
&&
-\frac{8}{3}D^2 (D/F)^3
< ( ( u_\mu, \, u_\nu)_\pm \, , \, {\lambda^a}^\dagger)_\pm  \, ( \lambda^b ,\,
u_\tau )_\pm >  \nonumber \\ 
&&
+\frac{32}{9}D^4 (D/F)
< ( {\lambda^a}^\dagger \, , \lambda^b )_\pm \, u_\mu > <   u_\nu  \,
u_\tau  >  \nonumber \\
&&
+\frac{32}{9}D^4 (D/F)
< ( {\lambda^a}^\dagger \, , \lambda^b )_\pm \, u_\tau > <   u_\mu  \,
u_\nu  >  \nonumber \\
&&
+\frac{32}{9}D^4 (D/F)
<  {\lambda^a}^\dagger \,  u_\mu > <  ( u_\nu  , \,
u_\tau )_\pm \, \lambda^b  >  \nonumber \\
&&
+\frac{32}{9}D^4 (D/F)
<  {\lambda^a}^\dagger \, ( u_\mu , \, u_\nu)_\pm  > <  
u_\tau \, \lambda^b  > \biggr\}
\eeqa

\vfill


\renewcommand{\arraystretch}{1.2}

\begin{table}
\caption{Counterterms and their $\beta$--functions as defined in
Eqs.(115,116)}
$$
\begin{tabular}{|r|c|c|} \hline
i  & $\tilde{O}_i^{ab}$ & $\beta_i$ \\ \hline
1 & $<{\lambda^a}^\dagger \,  u_\mu  > <  [i v \cdot \nabla , \, 
 u^\mu ] \, \lambda^b  >$ + h.c. & $ -4 D^4 /9$ \\
2 & $<{\lambda^a}^\dagger \, v \cdot u > <  [i v \cdot \nabla , \, 
v \cdot u ] \, \lambda^b >$ + h.c. & $4 D^4 /9$ \\
3 &  $  v_\rho \, \epsilon^{\rho\mu\nu\sigma} \, S_\sigma
<{\lambda^a}^\dagger \, \{ [ v \cdot \nabla , \, u_\mu ] , \, \{  
 u_\nu  , \, \lambda^b \} \} >$ + h.c. & $ -2D^2(D^2-F^2) / 9 $ \\
4 &  $  v_\rho \, \epsilon^{\rho\mu\nu\sigma} \, S_\sigma
<{\lambda^a}^\dagger \, [ [ v \cdot \nabla , \, u_\mu ] , \, [  
 u_\nu  , \, \lambda^b ] ] >$ + h.c. & $ 2D^2(D^2-F^2) / 9 $ \\
5  & $<{\lambda^a}^\dagger \{ (v \cdot u)^2 , [i v \cdot \nabla,
\lambda^b] \} >$ + h.c. & $3/4$ \\
6  & $<{\lambda^a}^\dagger [[ v \cdot u \, , \, [i v \cdot \nabla, v
\cdot u ]]\, ,\,  \lambda^b ] >$  & $3/4  + (D^4 - 10 D^2 F^2 + 9
F^4)/12 $ \\ 
7  & $<{\lambda^a}^\dagger [[  u_\mu \, , \, [i v \cdot \nabla, 
 u^\mu ]]\, ,\,  \lambda^b ] >$  & $  - (D^4 - 10 D^2 F^2 + 9 F^4)/12 $ \\
8  & $<{\lambda^a}^\dagger \{[ v \cdot u \, , \, [i v \cdot \nabla, v
\cdot u ]]\, ,\,  \lambda^b \} >$  & $ -4 D^3 F / 3 $ \\ 
9  & $<{\lambda^a}^\dagger \{[  u_\mu \, , \, [i v \cdot \nabla, 
 u^\mu ]]\, ,\,  \lambda^b \} >$  & $  4 D^3 F/ 3 $ \\
10  & $i<{\lambda^a}^\dagger [ [ u_\mu, [ \nabla^\mu ,  v \cdot u] ]
, \lambda^b ] >$  & $-(9+25D^2+45F^2)/72$ \\
11  & $i<{\lambda^a}^\dagger [[ [ \nabla^\mu ,u_\mu],  v \cdot u ]
, \lambda^b ] >$  & $-(9+25D^2+45F^2)/72$ \\
12  & $i<{\lambda^a}^\dagger \{ [ u_\mu, [ \nabla^\mu ,  v \cdot u ]]
, \lambda^b \} >$  & $- 5 DF / 4$ \\
13  & $i<{\lambda^a}^\dagger \{ [ [ \nabla^\mu ,u_\mu],  v \cdot u ]
, \lambda^b \} >$  & $- 5 D F / 4$ \\
14  & $  v_\rho \, \epsilon^{\rho\mu\nu\sigma} S_\sigma \,
<{\lambda^a}^\dagger  \, \{ \{  u_\mu , \, [ v \cdot \nabla , \,
u_\nu ] \} , \, \lambda^b \} >   
$   & $ -(15D^4 + 26 D^2F^2  - 9 F^4)/18 $ \\
15  & $  v_\rho \, \epsilon^{\rho\mu\nu\sigma} S_\sigma \,
<{\lambda^a}^\dagger  \, [ \{  u_\mu , \, [ v \cdot \nabla , \,
u_\nu ] \} , \, \lambda^b ] >   
$   & $  8 D^3F/9 $ \\
16  & $  v_\rho \, \epsilon^{\rho\mu\nu\sigma} S_\sigma \,
<{\lambda^a}^\dagger  \, \lambda^b > <  u_\mu \, [ v \cdot \nabla , \,
u_\nu ] >   
$   & $ (34D^4 + 108 D^2F^2  + 18F^4)/27 $ \\
17  & $v_\rho \, \epsilon^{\rho\mu\nu\sigma} \, S_\sigma
<{\lambda^a}^\dagger  u_\mu >  <  u_\nu   [v \cdot\nabla , 
 \lambda^b ] >$ + h.c. & $ 8 D^4/9 $ \\
18  & $v_\rho \, \epsilon^{\rho\mu\nu\sigma} \, S_\sigma
<{\lambda^a}^\dagger  u_\mu >  <   [v \cdot\nabla , \, u_\nu ]
 \, \lambda^b  >$ + h.c. & $ 2 ( 13 D^4 - 18 D^2 F^2 + 9 F^2 ) /27 $ \\
19  & $v_\rho \, \epsilon^{\rho\mu\nu\sigma} \, S_\sigma
<{\lambda^a}^\dagger \{ [ u_\mu, u_\nu],  [v \cdot\nabla , 
 \lambda^b ] \} >$ + h.c. & $ - 3 D F /2 -4 D^3 F/3  $ \\
20  & $v_\rho \, \epsilon^{\rho\mu\nu\sigma} \, S_\sigma
<{\lambda^a}^\dagger [ [ u_\mu, u_\nu],  [v \cdot\nabla , 
 \lambda^b ] ] >$ + h.c. & $- (5 D^2 + 9F^2)/12 $ \\
&& $ + (D^4 - 10 D^2F^2 +9F^4)/12 $ \\
21  & $i v_\rho \, \epsilon^{\rho\mu\nu\sigma} \, 
<{\lambda^a}^\dagger \, \lambda^b > <[ u_\mu, u_\nu] u_\sigma> 
$  & $- F(3 D^2 + F^2)/ 4$ \\
22  & $i v_\rho \, \epsilon^{\rho\mu\nu\sigma} \,
<{\lambda^a}^\dagger \, [ u_\mu, u_\nu]>< u_\sigma \,  \lambda^b > 
$+h.c.  & $ F( D^2 - F^2)/ 4$ \\
23  & $i v_\rho \, \epsilon^{\rho\mu\nu\sigma} \, 
<{\lambda^a}^\dagger \, \{ [ u_\mu, u_\nu],\{ u_\sigma \, , \lambda^b \}\} > 
$+h.c.  & $ 3F( D^2 - F^2)/ 32$ \\
24  & $i v_\rho \, \epsilon^{\rho\mu\nu\sigma} \, 
<{\lambda^a}^\dagger \, \{ [ u_\mu, u_\nu], [ u_\sigma \, , \lambda^b ] \} > 
$+h.c.  & $ -3D( D^2 - F^2)/ 32$ \\
25  & $i v_\rho \, \epsilon^{\rho\mu\nu\sigma} \, 
<{\lambda^a}^\dagger \, [ [ u_\mu, u_\nu], \{ u_\sigma \, , \lambda^b\}]> 
$+h.c.  & $ -D( 7D^2 +9 F^2)/ 96$ \\
26  & $i v_\rho \, \epsilon^{\rho\mu\nu\sigma} \, 
<{\lambda^a}^\dagger \, [ [ u_\mu, u_\nu], [ u_\sigma \, , \lambda^b ] ]> 
$+h.c.  & $ F( 7D^2 + 9F^2)/ 96$ \\ 
27 & $i<{\lambda^a}^\dagger [ v \cdot \nabla , [ v \cdot \nabla,
[ v \cdot \nabla, \lambda^b]]] >$  & $ -(20D^2+36F^2)/3$ \\
 \hline
\end{tabular}
$$
\end{table}

$\,$


\addtocounter{table}{-1}

\begin{table}
\caption{continued}
$$
\begin{tabular}{|r|c|c|} \hline
i  & $\tilde{O}_i^{ab} $ & $\beta_i$ \\ \hline
28  & $<{\lambda^a}^\dagger  u_\mu \, [i v \cdot \nabla,
\lambda^b] \, u^\mu  >$ + h.c. & $ -D^2   + 4 D^2(D^2-F^2)$ \\
29  & $<{\lambda^a}^\dagger \{  u^2 , [i v \cdot \nabla, 
 \lambda^b] \} >$ + h.c.  & $-(5D^2+9F^2)/8 + (15D^4+26 F^2 D^2- 9F^4) /4$ \\
30  & $<{\lambda^a}^\dagger [  u^2 , [i v \cdot \nabla, 
 \lambda^b] ] >$ + h.c.  & $-9 D F / 4 - 4 D^3 F  $ \\
31  & $<{\lambda^a}^\dagger \, [i \, v \cdot \nabla , \lambda^b]  >
<u \cdot u > $ + h.c. & $-3(D^2 + F^2) / 4 - (17D^4+54D^2F^2+9F^4) / 6$ \\
32  & $ <{\lambda^a}^\dagger  \, u_\mu><u^\mu [ i v \cdot \nabla, \lambda^b] >
$ + h.c. & $ 3 (D^2 - F^2) / 2  - (13D^4 -18D^2F^2 + 9 F^4) / 3$ \\
33  & $<{\lambda^a}^\dagger \{ \chi_+ , [i v \cdot \nabla, 
 \lambda^b] \} >$ + h.c.  & $3 (D^2 - 3F^2)/8$ \\
34  & $<{\lambda^a}^\dagger [ \chi_+ , [i v \cdot \nabla, 
 \lambda^b] ] >$ + h.c.  & $ - 5 D F / 4$ \\
35  & $<{\lambda^a}^\dagger \, [i v \cdot \nabla , \lambda^b]  >
<\chi_+ > $ + h.c. & $-(13D^2 +9 F^2) / 12$ \\
36  & $<[{\lambda^a}^\dagger , v \cdot \stackrel{\leftarrow}\nabla ] 
\{ S \cdot u, [ v \cdot \nabla ,\lambda^b] \} >
$  & $ 2 D (D^2 + 3 F^2) $ \\
37  & $<[{\lambda^a}^\dagger , v \cdot \stackrel{\leftarrow}\nabla ] 
[ S \cdot u, [ v \cdot \nabla ,\lambda^b] ] >
$  & $ -2 F (9 F^2 - 5 D^2) / 3 $ \\
38  & $<{\lambda^a}^\dagger \, \{ [ v \cdot \nabla , [v \cdot \nabla, 
 S \cdot u] ], \lambda^b \} > $  & $ -2D ( D^2 + 3 F^2) / 3$ \\
39  & $<{\lambda^a}^\dagger \, [ [ v \cdot \nabla , [v \cdot \nabla,
S \cdot u] ], \lambda^b ] > $  & $ -2F (5 D^2 - 9 F^2) / 9 $ \\
40  & $<{\lambda^a}^\dagger \, \{ S \cdot u ,  \lambda^b \} >
< \chi_+ > $  & $ D / 4 + 2D^3/9 + D(D^2-F^2)/4 $ \\
41  & $<{\lambda^a}^\dagger \, [ S \cdot u ,  \lambda^b ] >
< \chi_+ > $  & $ F / 4 +  2D^2F/9 - F(D^2-F^2)/4 $ \\
42  & $<{\lambda^a}^\dagger \,   \lambda^b  >
<  S \cdot u \, \chi_+ > $  & $ - D / 2 +  D(D^2+3F^2)/2 $ \\
43  & $<{\lambda^a}^\dagger \, \{ v \cdot u , \lambda^b \}  >
<v \cdot u \, S \cdot u > $  & $  - D - D(D^4 + 2D^2F^2 - 3F^4)/3  $ \\
44  & $<{\lambda^a}^\dagger \, [ v \cdot u , \lambda^b ]  >
<v \cdot u \, S \cdot u > $  & $  - F - F(-3D^4 + 2D^2F^2 + F^4)/3 $ \\
45  & $<{\lambda^a}^\dagger \, \{ S \cdot u ,v \cdot u\} > 
<v \cdot u \,  \lambda^b  > $ + h.c.   & $  D - D (D^2 - F^2)^2 / 3 $ \\
46  & $<{\lambda^a}^\dagger \, [ S \cdot u ,v \cdot u ] > 
<v \cdot u \,  \lambda^b  > $ + h.c.   & $  F - 16 D^4F/27 $ \\
47  & $<{\lambda^a}^\dagger \, \{ S \cdot u ,  \lambda^b \} >
< (v \cdot u)^2 > $  & $ D / 2 - D(D^4 + 2D^2F^2 - 3F^4)/6 - 16 D^5 / 27  $ \\
48  & $<{\lambda^a}^\dagger \, [ S \cdot u ,  \lambda^b ] >
< (v \cdot u)^2 > $  & $ F / 2 - F(-3D^4 + 2D^2F^2 + F^4)/6 + 16 D^4F
/27 $ \\
49  & $<{\lambda^a}^\dagger \,  S \cdot u  >
< (v \cdot u)^2 \,   \lambda^b  > $ + h.c.  & $ -D  - D (D^2 - F^2)^2
/ 3 - 16D^5/27$ \\
50  & $<{\lambda^a}^\dagger \,  \lambda^b  >
< (v \cdot u)^2 \,   S \cdot u  >  $   & $ D -D(D^4+10D^2F^2+5F^4) / 6 
- 16 D^5 / 27 $ \\
51  & $<{\lambda^a}^\dagger \, \{  v \cdot u \,   S \cdot u 
\, v \cdot u ,\lambda^b \} >  $   & $ 3 D/2  + D ( 5D^4 -30 D^2
F^2 +9 F^4)/18 $ \\
52  & $<{\lambda^a}^\dagger \, [  v \cdot u \,   S \cdot u 
\, v \cdot u ,\lambda^b ] >  $   & $ 3 F/2 + F ( -19D^4 - 6 D^2 F^2
+9 F^4)/18$ \\
53  & $<{\lambda^a}^\dagger \, \{  v \cdot u  ,   S \cdot u \} 
\, \lambda^b  v \cdot u  >  $ + crossed  & $ 2 D^3 ( D^2 -  F^2)/9 $ \\
54  & $<{\lambda^a}^\dagger \, \{  v \cdot u  ,   S \cdot u \} 
\, \lambda^b  v \cdot u  >  $ -- crossed  & $ 2 D^2F ( D^2 -  F^2)/9 $ \\
 \hline
\end{tabular}
$$
\end{table}

\addtocounter{table}{-1}

$\,$
\begin{table}
\caption{continued}
$$
\begin{tabular}{|r|c|c|} \hline
i  & $ \tilde{O}_i^{ab}$ & $\beta_i$ \\ \hline
55  & $<{\lambda^a}^\dagger \{  S \cdot u, \lambda^b \} >
<u^2> $  & $ D / 4 + D (D^2-F^2) / 4 $ \\
&& $ + D(D^4 + 2D^2F^2 - 3F^4)/6 + 16 D^5 / 27  $ \\
56  & $<{\lambda^a}^\dagger [  S \cdot u, \lambda^b ] >
<u^2> $  & $ F / 4 - F (D^2-F^2) / 4 $ \\
 && $ + F(-3D^4 + 2D^2F^2 + F^4)/6 - 16 D^4F / 27 $ \\
57  & $<{\lambda^a}^\dagger \,  \lambda^b  >
< S \cdot u \, u \cdot u> $  & $ -D / 2 + D (D^2 + 3F^2) / 2 $\\
& &
$ +D(D^4+10D^2F^2+5F^4) / 6 + 16 D^5 / 27 $ \\
58  & $<{\lambda^a}^\dagger \{ u_\mu ,  \lambda^b \}  >
< u^\mu \, S \cdot u > $  & $ D / 2 -  D (D^2 + 3F^2) / 2 $ \\
&& $  + D(D^4 + 2D^2F^2 - 3F^4)/3  $\\
59  & $<{\lambda^a}^\dagger [ u_\mu ,  \lambda^b ]  >
< u^\mu \, S \cdot u > $  & $ F / 2 - F(3D^2+F^2)/2 $ \\
&& $ + F(-3D^4 + 2D^2F^2 + F^4)/3 $ \\
60  & $<{\lambda^a}^\dagger  \{ S \cdot u, \{ (v \cdot u)^2  ,
\lambda^b \} \} >$ + h.c.  & $    D ( 29 D^4 + 28 D^2 F^2 - 9
F^4)/36$ \\
61  & $<{\lambda^a}^\dagger \{  S \cdot u, [ (v \cdot u)^2  ,
\lambda^b ] \} >$ + h.c.  & $ -3 F / 4 + F( -19 D^4 +12 D^2 F^2 - 9
F^4)/36 $ \\
62  & $<{\lambda^a}^\dagger [  S \cdot u, \{ (v \cdot u)^2  ,
\lambda^b \} ] >$ + h.c.  & $  F ( -23 D^4 +16 D^2 F^2 -9
F^4 )/36$ \\
63  & $<{\lambda^a}^\dagger [  S \cdot u, [ (v \cdot u)^2  ,
\lambda^b ] ]  >$ + h.c.  & $ -3 D / 4 + D ( 3 D^4  +16 D^2 F^2 -3
F^4)/12  $ \\
64  & $<{\lambda^a}^\dagger  v \cdot u \,  [i v \cdot \nabla,
\lambda^b] \,  v \cdot u  >$ + h.c. & $ -4 D^2(D^2-F^2)$ \\
65  & $<{\lambda^a}^\dagger \, v\cdot u><v\cdot u [i \, v \cdot
\nabla, \lambda^b] > 
$ + h.c. & $ 1 +  (13D^4 - 18 D^2F^2 + 9 F^2) / 3$ \\
66  & $<{\lambda^a}^\dagger [ (v \cdot u)^2 , [i v \cdot \nabla, 
 \lambda^b] ] >$ + h.c.  & $  4D^3F  $ \\
67  & $<{\lambda^a}^\dagger \{ (v \cdot  u)^2 , [i v \cdot \nabla, 
 \lambda^b] \} >$ + h.c.  & $ -(15D^4 + 26 F^2 D^2-9 F^4) / 4$ \\
68  & $<{\lambda^a}^\dagger \, [i \, v \cdot \nabla , \lambda^b]  >
<(v \cdot u)^2 > $ + h.c. & $ (3 + 17D^4+54D^2F^2+9F^4) / 6$ \\
69  & $<{\lambda^a}^\dagger \, u_\mu>< \{ u^\mu , \, S \cdot u \} 
 \lambda^b > $ +h.c. & $ -D (D^2 - F^2) / 2  + D (D^2 - F^2)^2 / 3 $ \\
70  & $<{\lambda^a}^\dagger \, u_\mu>< [ u^\mu , \, S \cdot u ] 
 \lambda^b > $ +h.c.  & $ -F (D^2 - F^2) / 2 + 16 D^4F / 27 $ \\
71  & $<{\lambda^a}^\dagger \,  S \cdot u >< u \cdot u \, \lambda^b  >
 $ +h.c.
   & $ D (D^2-F^2) / 2  + D (D^2 - F^2)^2/3 + 16 D^5/27 $ \\
72  & $<{\lambda^a}^\dagger  \{ S \cdot u, \{ u^2  ,
\lambda^b \} \} >$ + h.c.  & $ 3D(1 + D^2-F^2) /16 - D^3/3 $ \\
& &  $
- D( 29 D^4 + 28 D^2 F^2 - 9F^4)/36$ \\
73  & $<{\lambda^a}^\dagger \{  S \cdot u, [ u^2  ,
\lambda^b ] \} >$ + h.c.  & $  3F(1+ D^2-F^2) /16 $ \\
& & $ 
- F ( -19 D^4 + 12 D^2 F^2 - 9 F^4 )/36  $ \\
74  & $<{\lambda^a}^\dagger [  S \cdot u, \{  u^2  ,
\lambda^b \} ] >$ + h.c.  & $ - 3F(-1 + D^2-F^2) /16 + D^2F/3 $ \\
& &$
- F ( -23 D^4 +16  D^2 F^2 - 9F^4)/36  $ \\
75  & $<{\lambda^a}^\dagger [  S \cdot u, [ u^2  ,
\lambda^b ] ]  >$ + h.c.  & $ - 3D(-1 + D^2-F^2) /16 $ \\
 && $ 
- D ( 3 D^4 + 16 D^2 F^2 - 3F^4)/12 $ \\
 \hline
\end{tabular}
$$
\end{table}

$\,$

\addtocounter{table}{-1}

\begin{table}
\caption{continued}
$$
\begin{tabular}{|r|c|c|} \hline
i  & $\tilde{O}_i^{ab} $ & $\beta_i$ \\ \hline
76  & $<{\lambda^a}^\dagger \, \{  u_\mu \,   S \cdot u 
\,  u^\mu ,\lambda^b \} >  $   & $ -D ( 5D^4 -30 D^2
F^2 +9 F^4)/18 $ \\
77  & $<{\lambda^a}^\dagger \, [  u_\mu \,   S \cdot u 
\,  u^\mu ,\lambda^b ] >  $   & $ - F ( -19D^4 - 6 D^2 F^2
+9 F^4)/18$ \\
78  & $<{\lambda^a}^\dagger \, \{   u_\mu  ,   S \cdot u \} 
\, \lambda^b  u^\mu  >  $ + crossed  & $ 2D^3/3  - 2 D^3 ( D^2 -  F^2)/9 $ \\
79  & $<{\lambda^a}^\dagger \, \{   u_\mu  ,   S \cdot u \} 
\, \lambda^b  u^\mu  >  $ -- crossed & $   - 2 D^2F ( D^2 -  F^2)/9 $ \\
80  & $<{\lambda^a}^\dagger  \{ S \cdot u , \{ \chi_+ ,
\lambda^b \} \} >$ +h.c. & $  - D(-9 + 23D^2+9F^2)/48 $ \\
81  & $<{\lambda^a}^\dagger  \{ S \cdot u , [ \chi_+ ,
\lambda^b ] \} >$ +h.c. & $  -F(-9 + 7D^2+9F^2)/48 $ \\
82  & $<{\lambda^a}^\dagger  [ S \cdot u , \{ \chi_+ ,
\lambda^b \} ] >$ + h.c.  & $  - 3F(-1 + D^2-F^2)/16  $ \\
83  & $<{\lambda^a}^\dagger  [ S \cdot u, [ \chi_+ ,
\lambda^b ] ] >$ + h.c.  & $ - D(-9 + 9D^2+7F^2)/48 $ \\
 84  & $<{\lambda^a}^\dagger   S \cdot u  >  < \chi_+\lambda^b  > $ + h.c.
 & $ 4D^3/9 + D(D^2-F^2)/2 $ \\
 85  & $ i v_\rho \, \epsilon^{\rho\mu\nu\tau} \,
<{\lambda^a}^\dagger \, \lambda^b   >  < [ u_\mu ,\, u_\nu ]\, u_\tau  > 
$  & $  - F (5D^4+10D^2F^2+F^4) / 8 - 4 D^4 F / 9 $ \\
 86  & $ i \, v_\rho \, \epsilon^{\rho\mu\nu\tau} \,
<{\lambda^a}^\dagger  \,  \{ [u_\mu \, ,   u_\nu] , \, \{ u_\tau  \, , \lambda^b \} \}  > 
$ +h.c. &  $  F (71 D^4 +18D^2F^2 -9F^4)/96 $    \\
 87  & $ i \, v_\rho \, \epsilon^{\rho\mu\nu\tau} \,
<{\lambda^a}^\dagger  \,  \{ [u_\mu \, ,   u_\nu] , \, [ u_\tau  \, , \lambda^b ] \}  > 
$ + h.c. &   $  D (-9D^4 - 30 D^2F^2 -9F^4)/96 $ \\
 88  & $ i \, v_\rho \, \epsilon^{\rho\mu\nu\tau} \,
<{\lambda^a}^\dagger  \,  [ [u_\mu \, ,   u_\nu] , \, \{ u_\tau  \, , \lambda^b  \} ] > 
$ +h.c. &   $  D (-D^4 -38 D^2F^2 -9F^4)/96 $  \\
 89  & $ i \, v_\rho \, \epsilon^{\rho\mu\nu\tau} \,
<{\lambda^a}^\dagger  \,  [ [u_\mu \, ,   u_\nu] , \, [ u_\tau  \, , \lambda^b ] ]  > 
$ +h.c & $  F (31 D^4 +58D^2F^2-9F^4)/96 $   \\
 90  & $ i\, v_\rho \, \epsilon^{\rho\mu\nu\tau} \,
<{\lambda^a}^\dagger  \, [ u_\mu \, , \, u_\nu ] >  <  
u_\tau \,  \lambda^b >  
$ + h.c.  & $  -(D^2-F^2)^2F/4 -4D^4F/9$ \\  
91  & $ v_\rho \, \epsilon^{\rho\mu\nu\sigma} \,
<{\lambda^a}^\dagger \, \lambda^b > < F_{\mu \nu}^+ \, u_\sigma> 
$  & $- F(3 D^2 + F^2)/ 2$ \\
92 & $v_\rho \, \epsilon^{\rho\mu\nu\sigma} \, 
<{\lambda^a}^\dagger \,  u_\sigma >< F_{\mu \nu}^+ \, \lambda^b > 
$+h.c.  & $ F( D^2 - F^2)/ 2$ \\
93  & $ v_\rho \, \epsilon^{\rho\mu\nu\sigma} \, 
<{\lambda^a}^\dagger \, \{  u_\sigma, \{ F_{\mu \nu}^+ \,, 
\lambda^b \}\} > $+h.c.  & $ 3F( D^2 - F^2)/ 16$ \\
94  & $ v_\rho \, \epsilon^{\rho\mu\nu\sigma} \, 
<{\lambda^a}^\dagger \, \{  u_\sigma, [F_{\mu \nu}^+ \,, 
\lambda^b ] \} > $+h.c.  & $ -D( 7D^2 + 9 F^2)/ 48$ \\
95  & $ v_\rho \, \epsilon^{\rho\mu\nu\sigma} \, 
<{\lambda^a}^\dagger \, [  u_\sigma,  \{ F_{\mu \nu}^+ \,, 
\lambda^b \} ] > $+h.c.  & $ -3D( D^2 - F^2)/ 16$ \\
96  & $ v_\rho \, \epsilon^{\rho\mu\nu\sigma} \, 
<{\lambda^a}^\dagger \, [  u_\sigma,  [ F_{\mu \nu}^+ \,, 
\lambda^b ] ] > $+h.c.  & $ F ( 7D^2 + 9F^2)/ 48$ \\
97  & $i v_\rho \, \epsilon^{\rho\mu\nu\sigma} \, S_\sigma
<{\lambda^a}^\dagger \, \{ F_{\mu \nu}^+ , [ v \cdot \nabla , 
\lambda^b ] \} > $+h.c.  & $ 3 D F $ \\
98  & $i v_\rho \, \epsilon^{\rho\mu\nu\sigma} \, S_\sigma
<{\lambda^a}^\dagger \, [ F_{\mu \nu}^+ , [ v \cdot \nabla , 
\lambda^b ] ] > $+h.c.  & $ (5 D^2 + 9 F^2)/6 $ \\
99  & $i \, S^\mu \, v^\nu \,
<{\lambda^a}^\dagger \, \{[ F_{\mu \nu}^+ ,  v \cdot u] , 
\lambda^b \} > $  & $ 3 D /2 $ \\
100  & $i \, S^\mu \, v^\nu \,
<{\lambda^a}^\dagger \, [ [F_{\mu \nu}^+ , v \cdot u] , 
\lambda^b ] > $  & $ 3 F /2 $ \\
101  & $<{\lambda^a}^\dagger \, [[ \nabla^\mu ,F_{\mu \nu}^+ v^\nu] , 
\lambda^b ] > $  & $-(9+25D^2+45F^2)/36 $ \\
102  & $<{\lambda^a}^\dagger \, \{[ \nabla^\mu ,F_{\mu \nu}^+ v^\nu] , 
\lambda^b \} > $  & $- 5 D F / 2 $ \\
\hline
\end{tabular}
$$
\end{table}




\section*{Figure Captions}

\begin{enumerate}

\item[Fig.~1] Contributions to the one--loop generating functional 
  ($\Sigma_1,\Sigma_2,\Sigma_3$) and
  the tree level mesonic generating functional at order $\hbar$,
  $\Sigma_4$. The solid (dashed) double lines denote the baryon
  (meson) propagator in the presence of external fields,
  respectively. The circle--cross in $\Sigma_4$ denotes the counter
  terms from ${\cal L}_M^{(4)}$. The contributions $\Sigma_{1,2}$ are
  called irreducible where as $\Sigma_{3,4}$ are reducible.

\item[Fig.~2] Some physical processes encoded in the irreducible
  generating functional. For the self--energy ($\Sigma_1$) and the
  tadpole graph ($\Sigma_2$), typical contributions from single and
  double pion photo/electroproduction are shown. Solid, dashed and
  wiggly lines denote baryons, pions and photons, in order.

\item[Fig.~3] Feynman graphs which lead to the divergence in the
   P--wave multipole $P_3$ for $\gamma p \to \pi^0 p$ as discussed
  in the text. $Y$ and $Y'$ can either be the $\Lambda$ or the
  $\Sigma^0$. Crossed graphs are not shown.

\end{enumerate}

\hbox{}\vspace{3.8cm}
\centerline{
\epsfysize=2.5in
\epsffile{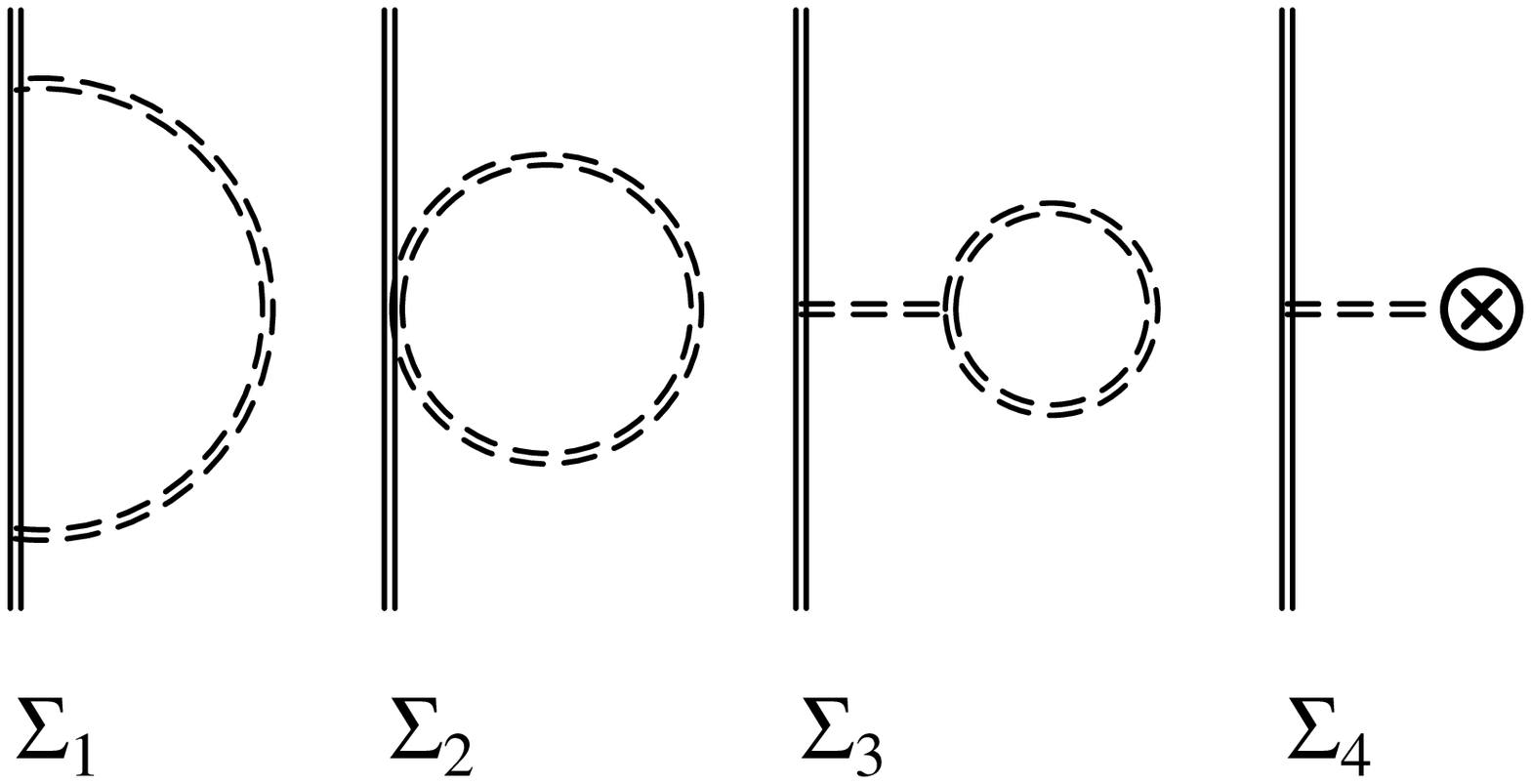}
}
\bigskip\bigskip\bigskip
\centerline{\Large Figure 1}

\vfill 


\hbox{}\vspace{0.3cm}

\centerline{
\epsfysize=3in
\epsffile{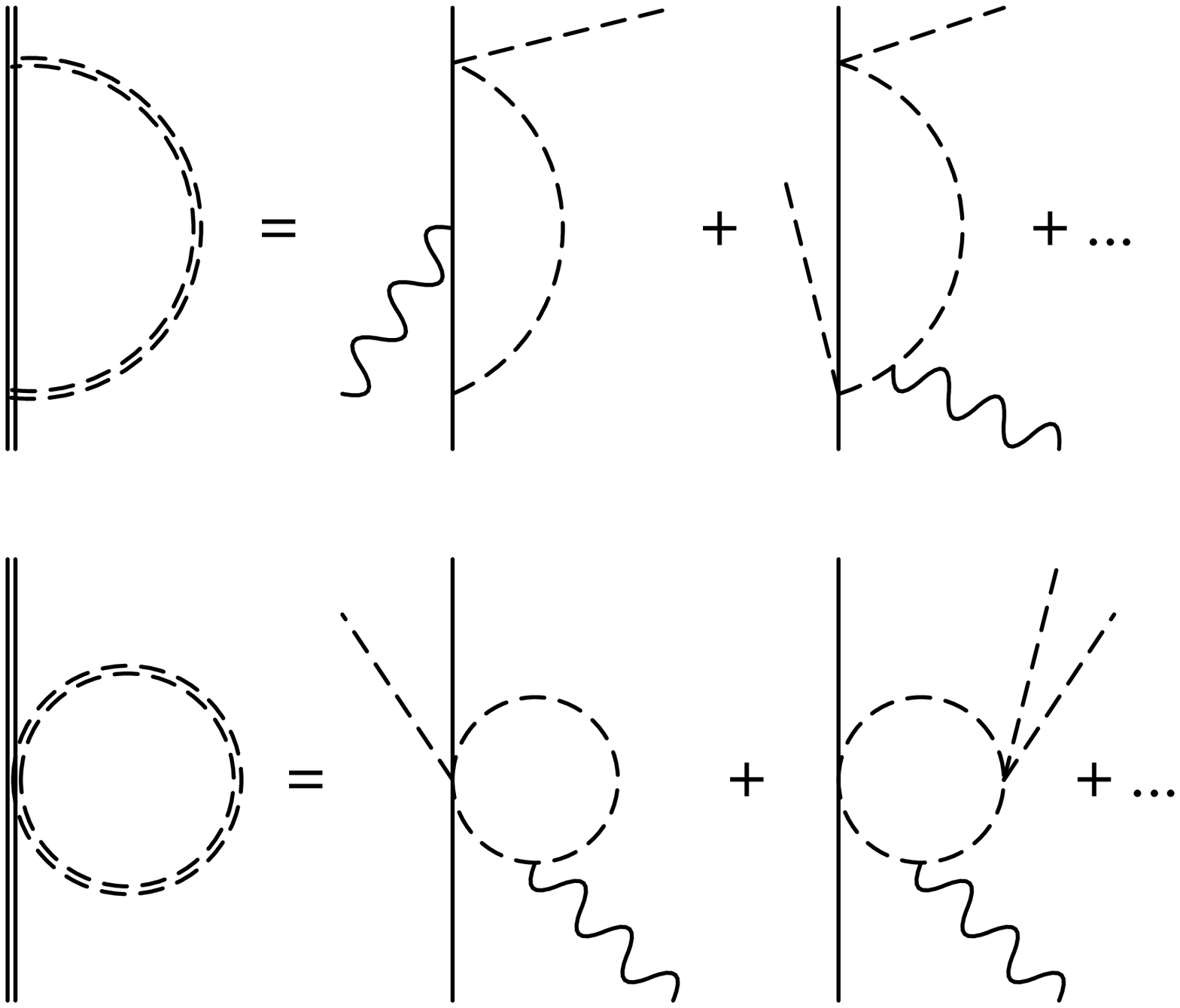}
}
\bigskip\bigskip
\centerline{\Large Figure 2}


\bigskip\bigskip\bigskip\bigskip

\centerline{
\epsfysize=3.5in
\epsffile{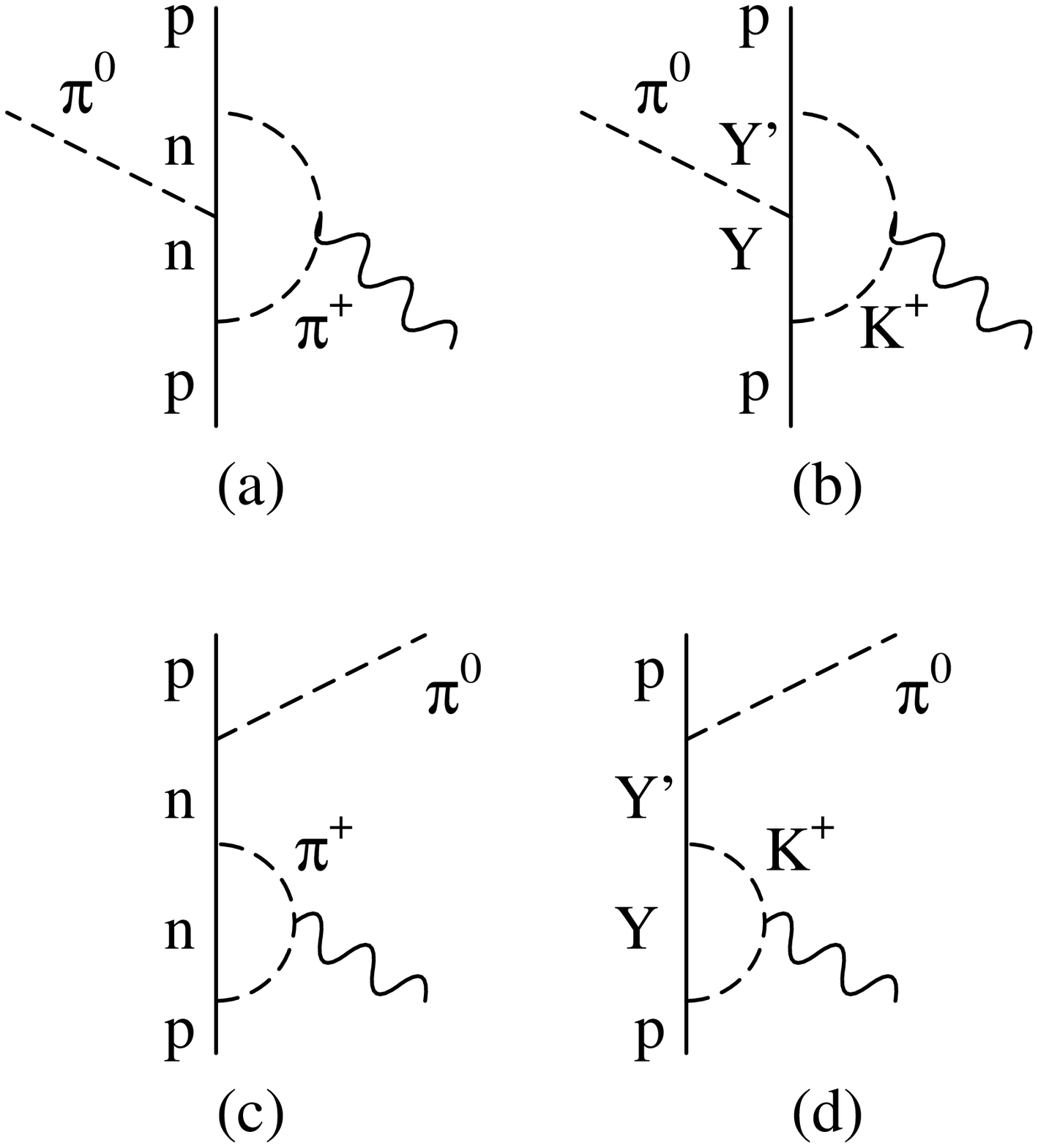}
}
\bigskip\bigskip
\centerline{\Large Figure 3}



\begin{thebibliography}{99}

\frenchspacing

\bibitem{wein68} S. Weinberg, Phys. Rev. {\bf 166} (1968) 1568

\bibitem{ccwz} S. Coleman, J. Wess and B. Zumino, Phys. Rev. {\bf 177}
  (1969) 2239

%

\bibitem{wein79} S. Weinberg, Physica {\bf 96A} (1979) 327

\bibitem{gss} J. Gasser, M.E. Sainio and A. Svarc, Nucl. Phys. {\bf
    B307} (1988) 779

\bibitem{jm} E. Jenkins and A.V. Manohar, Phys. Lett. {\bf B255} (1991) 558

\bibitem{bkkm}
V. Bernard, N. Kaiser, J. Kambor and Ulf-G. Mei\ss ner, Nucl. Phys.
{\bf B388} (1992) 315

\bibitem{ecker} G. Ecker, Phys. Lett.  {\bf B336} (1994) 508

\bibitem{bkmpipin}
V. Bernard, N. Kaiser and Ulf-G. Mei\ss ner, Nucl. Phys.
{\bf B457} (1995) 147

\bibitem{eckmoj} G. Ecker and M. Mojzis, Phys. Lett. {\bf B365} (1996) 312

\bibitem{ulfrev} Ulf-G. Mei{\ss}ner, Rep. Prog. Phys. {\bf 56} (1993) 903

\bibitem{bkmrev} V. Bernard, N. Kaiser and Ulf-G. Mei{\ss}ner,
Int. J. Mod. Phys. {\bf E4} (1995) 193

\bibitem{eckerrev} G. Ecker, Prog. Nucl. Part. Phys. {\bf 35} (1995) 1

\bibitem{rho} C.H. Lee et al., Nucl. Phys. {\bf A585} (1995) 401

\bibitem{sven} S. Steininger, Diploma thesis, University of Bonn,  
TK-96-26

\bibitem{bss} M.N. Butler, M. Savage and R. Springer, Nucl. Phys. {\bf
B399} (1993) 69

\bibitem{krause} A. Krause, Helv. Phys. Acta {\bf 63} (1990) 3


\bibitem{gl85} J. Gasser and H. Leutwyler, Nucl. Phys. {\bf B250} (1985) 465

\bibitem{mannel} T. Mannel, W. Roberts and Z. Ryzak, Nucl. Phys. {\bf
    B368} (1992) 204
 
\bibitem{eckerc} G. Ecker, Czech. J. Phys. {\bf 44} (1994) 405

\bibitem{gilkey} D.  Gilkey, ``Index Theorems and the Heat Equation'',
Publish or Perish, Berkley, 1975

\bibitem{ball} R.D. Ball, Phys. Rep. {\bf 182} (1989) 1

\bibitem{jacko} I. Jack and H. Osborn, Nucl. Phys. {\bf B207} (1982) 474

\bibitem{bkmz} V. Bernard, N. Kaiser and Ulf-G. Mei\ss ner, Z. Phys.
{\bf C70} (1996) 483

\bibitem{sch} H. Fearing and S. Scherer, Phys. Rev. {\bf D53} (1996) 315 


\end{thebibliography}
\end{document}